\documentclass[a4paper,journal,twocolumn]{IEEEtran}

\ifCLASSINFOpdf
\else
\fi

\usepackage[cmex10]{amsmath}
\usepackage{amssymb}
\usepackage{mathtools}
\usepackage{epsfig}
\usepackage{epstopdf}
\usepackage{subfigure}
\usepackage{booktabs}
\usepackage{multirow}
\usepackage{blindtext}
\usepackage{algorithm}
\usepackage{algorithmic}
\usepackage{color}
\usepackage{amsmath}
\usepackage{graphicx}
\usepackage{pgfplots}
\pgfmathdeclarefunction{Krone}{1}{\pgfmathparse{or(abs(#1) > 0,0)}}
\usepackage{bbm}

\usepackage[noadjust]{cite}
\newsavebox{\ieeealgbox}

\usepackage{array}

\usepackage{mdwmath}
\usepackage{mdwtab}

\usepackage{stfloats}
\fnbelowfloat

\newcommand{\Ab}{\mathbf{A}}
\newcommand{\ab}{\mathbf{a}}

\newcommand{\Ib}{\mathbf{I}}

\newcommand{\bb}{\mathbf{b}}
\newcommand{\Wb}{\mathbf{W}}
\newcommand{\Xb}{\mathbf{X}}
\newcommand{\xb}{\mathbf{x}}
\newcommand{\xbs}{\xb^*}
\newcommand{\xbh}{\widehat{\xb}}
\newcommand{\xbt}{\widetilde{\xb}}

\newcommand{\ub}{\mathbf{u}}
\newcommand{\vb}{\mathbf{v}}
\newcommand{\hb}{\mathbf{h}}

\newcommand{\yb}{\mathbf{y}}
\newcommand{\ybt}{\widetilde{\yb}}
\newcommand{\ybs}{\yb^*}
\newcommand{\zb}{\mathbf{z}}
\newcommand{\wb}{\mathbf{w}}
\newcommand{\Yb}{\mathbf{Y}}
\newcommand{\Zb}{\mathbf{Z}}
\newcommand{\Zerb}{\mathbf{0}}

\newcommand{\xbsi}{\xb_{\sigma}}
\newcommand{\Rbb}{\mathbb{R}}

\newcommand{\UDelmOne}{\lceil \Delta-1 \rceil}

\newcommand{\SIGb}{\boldsymbol{\Sigma}}

\newcommand{\fsig}{f_{\sigma}}
\newcommand{\Fsig}{F_{\sigma}}

\newcommand{\summ}{\sum_{i=1}^{m}}
\newcommand{\sumn}{\sum_{i=1}^{n}}

\newcommand{\SNR}{\text{SNR}_{\text{rec}}}
\newcommand{\MSNR}{\text{MSNR}_{\text{rec}}}

\DeclareMathOperator{\diag}{diag}

\DeclareMathOperator*{\argmin}{argmin}

\DeclareMathOperator{\nullS}{null}

\DeclareMathOperator{\sign}{sign}
\DeclareMathOperator{\median}{median}

\newcommand{\NullDefmZ}{\nullS(\Ab)\setminus \{\Zerb\}}

\definecolor{brown}{rgb}{0.5, 0.3, 0.1}
\definecolor{purple}{rgb}{0.625, 0.125, 0.905}
\definecolor{orange}{rgb}{1, 0.27, 0}
\definecolor{DarkGreen}{rgb}{0.0, 0.5, 0.0}

\def\ROne_Col{black}
\def\RTwo_Col{black}
\def\MinEdt{black}

\begin{document}
\title{{\color{\RTwo_Col}Successive Concave Sparsity Approximation for Compressed Sensing}}
%
%
%

\author{Mohammadreza~Malek-Mohammadi\textsuperscript{*},~\IEEEmembership{Member,~IEEE}, Ali~Koochakzadeh, Massoud~Babaie-Zadeh,~\IEEEmembership{Senior Member,~IEEE},~Magnus~Jansson,~\IEEEmembership{Member,~IEEE},~and Cristian R. Rojas,~\IEEEmembership{Member,~IEEE}
\thanks{{\color{\MinEdt}This work was supported by the Swedish Research Council under contract 621-2011-5847 and the Swedish Strategic Research Area ICT-TNG program.}}
\thanks{{\color{\MinEdt}M.~Malek-Mohammadi, M.~Jansson and C.~R.~Rojas are with the ACCESS Linnaeus Centre, KTH (Royal Institute of Technology), Stockholm, 10044, Sweden (e-mail: \{mohamma,janssonm,crro\}@kth.se). A.~Koochakzadeh is with the Department of Electrical and Computer Engineering, University of Maryland, College Park, MD 20742 USA (email: alik@umd.edu). M. Babaie-Zadeh is with the Department of Electrical Engineering, Sharif University of Technology, Tehran 1458889694, Iran (e-mail: mbzadeh@yahoo.com).}}}
\maketitle

\begin{abstract}
In this paper, based on a successively accuracy-increasing approximation of the $\ell_0$ norm, we propose a new algorithm for recovery of sparse vectors from underdetermined measurements. The approximations are realized with a certain class of concave functions that aggressively induce sparsity and their closeness to the $\ell_0$ norm can be controlled. We prove that the series of the approximations asymptotically coincides with the $\ell_1$ and $\ell_0$ norms when the approximation accuracy changes from the worst fitting to the best fitting. When measurements are noise-free, an optimization scheme is proposed which leads to a number of weighted $\ell_1$ minimization programs, whereas, in the presence of noise, we propose two iterative thresholding methods that are computationally appealing. A convergence guarantee for the iterative thresholding method is provided, and, for a particular function in the class of the approximating functions, we derive the closed-form thresholding operator. We further present some theoretical analyses via the restricted isometry, null space, and spherical section properties. Our extensive numerical simulations indicate that the proposed algorithm closely follows the performance of the oracle estimator for a range of sparsity levels wider than those of the state-of-the-art algorithms.
\end{abstract}


%

\begin{center} \bfseries EDICS Category: DSP-CPSL, DSP-RECO, SAM-CSSM, OPT-NCVX \end{center}

\section{Introduction}
\IEEEPARstart{D}{uring} the last decade with the appearance of theoretical and experimental results showing the possibility of recovering sparse signals from undersampled measurements, there has been an explosion of applications gaining from the reduction in the sampling rate. In fact, any field of science or engineering, where sampling a signal is part of the processing task, can potentially benefit from the line of research ongoing in the domain of compressed sensing (CS). This can be witnessed with applications in sciences such as quantum state tomography in physics \cite{GrosLFBE10}, imaging in astrophysics \cite{WiauJPSV09}, bacterial community reconstruction in biology \cite{KoslFR13}, and genotyping in genetics \cite{ErliGBHM10}. In engineering, the number of applications is tremendous and includes magnetic resonance imaging \cite{LustDP07}, sampling of analog signals \cite{Elda09}, array signal processing \cite{Bili11}, and radar \cite{HermS09}, to name a few.

A general formulation for the sampling in the compressed fashion is
\begin{equation} \label{GenModel}
\bb = \Ab \xbt + \wb,
\end{equation}
in which $\bb \in \Rbb^{n}$ is the vector of measurements, $\Ab \in \Rbb^{n \times m}~(n < m)$ is the sensing matrix, $\xbt \in \Rbb^{m}$ is the unknown sparse vector with $s \ll m$ nonzero elements, and $\wb$ is the probable additive noise. In a nutshell, given $\bb$ and knowing $\Ab$, the goal in recovery of sparse vector from compressed measurements is to accurately estimate $\xbt$. As \eqref{GenModel} is underdetermined, recovery of $\xbt$ from $\bb$ is ill-posed unless we know \emph{a priori} that $\xbt$ resides in a low-dimensional space. Sparsity enters the game, and the sparsest solution which is consistent to the measurements is sought via
\begin{equation} \label{l0minNoisy}
\min_{\xb} \|\xb\|_0  \quad \text{subject to} \quad \| \Ab\xb - \bb\| \leq \epsilon
\end{equation}
in which $\| \xb \|_0$ designates the so-called $\ell_0$ norm of $\xb$ defined as its number of nonzero entries, $\| \cdot \|$ stands for the $\ell_2$ norm, and $\epsilon \geq \| \wb \|$ is some constant.

Under certain circumstances, the solution to \eqref{l0minNoisy} is close to $\xbt$ in \eqref{GenModel} \cite{DonoET06,BabaJ10}, and, particularly, when $\wb = \Zerb$ (i.e., measurements are noise-free),
\begin{equation} \label{l0min}
\min_{\xb} \|\xb\|_0  \quad \text{subject to} \quad \Ab\xb=\bb
\end{equation}
exactly recovers $\xbt$ \cite{DonoET06}. Programs \eqref{l0minNoisy} and \eqref{l0min} are generally NP-hard \cite{Elad10}; however, their convex relaxations,
\begin{equation} \label{l1minNoisy}
\min_{\xb} \|\xb\|_1  \quad \text{subject to} \quad \| \Ab\xb - \bb\| \leq \epsilon
\end{equation}
and
\begin{equation} \label{l1min}
\min_{\xb} \|\xb\|_1  \quad \text{subject to} \quad \Ab\xb=\bb,
\end{equation}
can be considered instead. Convexification makes the recovery tractable at the cost of increasing the sampling rate needed for stable recovery along with increase in the reconstruction error \cite{DonoET06}. A hot topic of research is, therefore, to propose recovery algorithms to push the sampling rate and reconstruction error as much as practically possible toward the intrinsic bounds of $\ell_0$ minimization. To this end, several nonconvex alternatives for \eqref{l0min} and \eqref{l0minNoisy} have been proposed in the literature. The list for the noise-free recovery is already rich (cf. \cite{CandWB08,Char07,MohiBJ09,FoucL09,Rang11}), yet, in the noisy recovery, there is still much room for improvement.

\subsection{{\color{\MinEdt}Contribution}}

In this paper, we propose a new algorithm for recovery of sparse vectors. Although, as will be shown in the numerical simulations, the proposed algorithm outperforms some of the state-of-the-art algorithms in the noise-free recovery, the main goal of introducing this method is to provide an effective algorithm for the more challenging problem of recovery from compressed noisy measurements. The core idea of the proposed method is to closely approximate the $\ell_0$ norm with a certain class of concave functions and then to minimize this approximation subject to the constraints. However, in contrast to the ideas of \cite{Char07} and \cite{CandWB08} where a fixed approximation is used, we use a series of approximations in which the accuracy of the approximations improves as the algorithm proceeds. 
The proposed algorithm is, hence, called SCSA standing for successive concave sparsity approximation.

In the noise-free case, our proposed algorithm involves solving some weighted $\ell_1$ minimization programs which is computationally demanding. Nonetheless, in the noisy case, we utilize an iterative thresholding method \cite{CombP11} to decrease the complexity and derive a closed-form solution for the thresholding operator. In addition, we theoretically characterize the conditions under which the proposed iterative thresholding method converges.

On the theoretical side, the SCSA algorithm is supported by some guarantees derived from the restricted isometry \cite{Cand08}, null space \cite{GribN07}, and spherical section \cite{KashT07} properties. On the numerical side, extensive empirical evaluations in the noisy case show that the SCSA algorithm significantly and consistently outperforms $\ell_1$ minimization as well as some other {\color{\MinEdt}nonconvex} methods in terms of reconstruction error, whereas the execution time is at most three times longer than that of {\color{\MinEdt}one of} the fastest algorithms in the comparison. Furthermore, we show that the SCSA algorithm 
closely follows the oracle estimator \cite{CandT07} for a broad range of sparsity levels.

\subsection{{\color{\MinEdt}Connections to Previous Work}}

{\color{\ROne_Col}The idea of the SCSA algorithm is borrowed from \cite{MaleBS14}} that proposes a method for low-rank matrix recovery. In this work, we apply the same idea to the sparse recovery problem and propose an efficient optimization method for the noisy case which is not considered in \cite{MaleBS14}. Also, the underlying idea of SCSA is somehow connected to the ideas of the SL0 algorithm \cite{MohiBJ09} and the algorithm of \cite{TrzaM09}. The fundamental difference with the SL0 algorithm is as follows. Let $\sigma$ denote the parameter that reflects the accuracy of $\ell_0$ norm approximation for SCSA, SL0, and the method of \cite{TrzaM09}, where a smaller $\sigma$ corresponds to better accuracy. For SCSA, we use a class of subadditive functions which enables us to analytically prove that, under some conditions, minimizing the $\ell_0$ norm approximation for any $\sigma > 0$ leads to exact or accurate recovery in the noiseless or noisy cases. However, there is not such a guarantee for SL0 except for the asymptotic case of $\sigma \to 0$. This may also justify the performance improvement over SL0 demonstrated in our numerical simulations. Additionally, as the approximating functions differ, completely dissimilar optimization methods are exploited. In comparison to \cite{TrzaM09}, the main distinctions are summarized as:
\begin{itemize}
  \item The method of \cite{TrzaM09} has been proposed for the noise-free case and has been specialized for ``reconstruction of sparse images,'' while SCSA applies to the noisy case as well and is designed for the general framework of CS.
  \item In this work, theoretical analyses for the asymptotic cases of $\sigma \to \infty$ and $\sigma \to 0$ are provided.
  \item Here, convergence analysis for the proposed optimization methods are also given.
  \item \cite{TrzaM09} solves the associated optimization problem by smoothing the approximating functions to make it simpler, while SCSA introduces no smoothing.
  \item Finally, in this paper, the proposed initialization is theoretically {\color{\RTwo_Col}motivated}, while \cite{TrzaM09} intuitively justifies its different initialization.
\end{itemize}

\subsection{{\color{\MinEdt}Notations and Outline}}

$\lceil  x \rceil$ designates the smallest integer greater than or equal to $x$. $\sign(x) = |x| / x$ for $x \in \Rbb \setminus \{0\}$, and $\sign(0) = 0$. The $\sign(\cdot)$, $\max(\cdot,\cdot)$ and $\min(\cdot,\cdot)$ functions act component-wise on vector inputs. For a vector $\xb$, $\| \xb \|_1$ and $\| \xb \|$ denote the $\ell_1$ and $\ell_2$ norms, respectively, $\| \xb \|_0$ denotes the so-called $\ell_0$ norm, and $x_i$ represents the $i$th element. For vectors $\xb$  and $\yb$, $\xb \ge \yb$ means that $\forall i, x_i \geq y_i$, $\xb \odot \yb$ indicates component-wise multiplication, and $\langle \xb, \yb \rangle = \xb^T \yb$ denotes the inner product. $\Xb ^\dagger$ denotes the Moore-Penrose pseudoinverse of the matrix $\Xb$. For symmetric matrices $\Yb,\Zb$, $\Yb \succeq \Zb$ means $\Yb - \Zb$ is positive semidefinite. For a positive semidefinite matrix $\Xb$, $\lambda_{max}(\Xb)$ and $\lambda_{min}(\Xb)$ denote the largest and smallest eigenvalues. $\nullS(\Ab)$ represents the null space of the matrix $\Ab$. $\ensuremath{N(\Zerb,\SIGb)}$ designates the multivariate Gaussian distribution with mean $\Zerb$ and covariance $\SIGb$.

The rest of this paper is structured as follows. In Section~\ref{sec:algomain}, the main idea of the SCSA algorithm is introduced, and Section~\ref{sec:algoimp} explains the optimization methods used in the noiseless and noisy cases. In Section~\ref{sec:PerAn}, theoretical analyses are presented, and, in Section~\ref{sec:NumExp}, the performance of the SCSA method is evaluated and compared against some state-of-the-art algorithms by numerical simulations. Section~\ref{sec:Con} concludes the paper.

\section{Main idea of the SCSA Algorithm} \label{sec:algomain}
\subsection{Motivation}
The problem of finding the sparsest solution of $\Ab\xb=\bb$ or the sparsest vector in the set $\{ \xb~|~ \| \Ab\xb - \bb\| \leq \epsilon\}$ can be interpreted as the task of approximating the Kronecker delta function (called herein the delta function for brevity). Let
\begin{equation*}
\delta(x) = \left\{
\begin{IEEEeqnarraybox}[][c]{l?s}
\IEEEstrut
1 & if $x = 0$ \\
0 & otherwise
\IEEEstrut
\end{IEEEeqnarraybox}
\right.
\end{equation*}
denote the delta function, then the $\ell_0$ norm of a vector $\xb$ is equal to $\| \xb\|_0 = \summ [1 - \delta(x_i)]$ and can be approximated by $\summ f(x_i)$ in which $f:\Rbb \to \Rbb$ acts as a delta approximating (DA) function.\footnote{{\color{\MinEdt}Clearly, $f(x)$ is approximating $1-\delta(x)$, not the delta function. Nevertheless, we call it DA function for the sake of easy referral.}} Replacing the $\ell_0$ norm with the above approximation, the next step is to find a point in the feasible set that minimizes $\summ f(x_i)$. To have a numerically tractable optimization problem, one needs to find suitable DA functions with some appropriate properties like convexity or continuity. Based on this formulation, $f(x) = |x|$ promotes sparsity with the $\ell_1$ norm, and $f(x) = |x|^p, 0 < p < 1,$ gives rise to $\ell_p$ quasi-norm minimization. Another well-known example is $\log(x)$ which leads to reweighted $\ell_1$ minimization \cite{CandWB08}.

Though $\ell_1$ minimization enjoys convexity, a number of theoretical and numerical analyses show the superiority of other DA choices. For instance, \cite{CharS08} and \cite{WuC13} show that, at least for some $p$'s in $(0,1)$ and under smaller thresholds on the restricted isometry constant (RIC) \cite{Cand08}, 
globally minimizing the $\ell_p$ quasi-norm subject to the linear equations recovers sparse vectors uniquely. Also, \cite{Zhan10,ZhanZ12} prove that, in noisy CS and under some mild conditions, even locally minimizing the $\ell_p$ quasi-norm via a certain optimization scheme leads to recovery of a solution with less error. 
On the numerical side, \cite{Char07, Char07b, CharS08} give some numerical evaluations demonstrating the superiority of $\ell_p$ quasi-norm minimization. Moreover, \cite{ShenFL13,CandWB08,Need09} analyze and compare the theoretical and/or numerical performance of reweighted $\ell_1$ minimization to $\ell_1$ minimization showing considerable improvements.

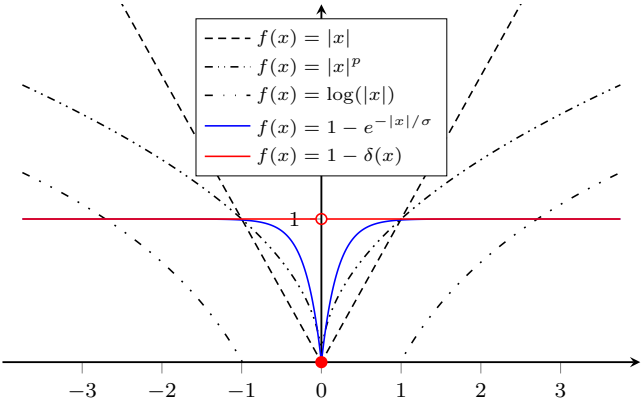
\begin{figure}[tb]
\centering
\pgfplotsset{every axis/.append style={font = \footnotesize, line width=0.7pt}}
\begin{tikzpicture}
	\begin{axis}[
	width = 0.55\textwidth,
	height = 180pt,
	ymin = 0, ymax = 2.5,
	xmin = -4, xmax = 4,
	ytick={1,2},
	xtick={-3,-2,-1,0,1,2,3},
	axis x line=bottom,
	axis y line=center,
	scaled ticks = false,
	tick align=outside,
	legend cell align = left,
    legend style={font=\scriptsize,at={(0.5,0.74)},anchor=center,line width = 0.3},
    legend entries ={$f(x) = |x|$,$f(x) = |x|^p$,$f(x) = \log(|x|)$,$f(x) = 1 - e^{-|x|/\sigma}$,$f(x) =1 - \delta(x)$}
	]
	\addplot[black,densely dashed,domain=-3.75:3.75,samples = 1000,line width = 0.6] (\x,{abs(\x)});
	\addplot[black,dashdotdotted,domain=-3.75:3.75,samples = 1000,line width = 0.6] (\x,{sqrt(abs(\x))});
	\addplot[black,loosely dashdotdotted,domain=-3.75:3.75,samples = 1000,line width = 0.6] (\x,{ln(abs(\x)+0.01)});
	\addplot[blue,domain=-3.75:3.75,samples = 1000,line width = 0.6] (\x,{1 - exp(-abs(\x)/0.2)});
	\addplot[red,domain=-3.75:-0.07,samples = 500,line width = 0.6] (\x,{Krone(\x)});
	\addplot[red,domain=0.07:3.75,samples = 500,line width = 0.6] (\x,{Krone(\x)});
	\addplot+[red,mark=o,line width = 0.6] coordinates {(0,1)};
	\addplot[red,mark color=red,mark=*,line width = 0.6] coordinates {(0,0)};

	\end{axis}
\end{tikzpicture}
\caption{Some algorithms for recovery of sparse vectors implicitly or explicitly use different functions to approximate the delta function as $\| \xb\|_0$ equals $\sumn 1 - \delta(x_i)$. Some of them are plotted in this figure, and among them, $\fsig(x) = 1 - e^{-|x|/\sigma}$ more closely matches $1 - \delta(x)$ for a small $\sigma$.} \label{fig:FigFunc}
\end{figure}

With this background, one may think whether better approximations of the delta function lead to higher performance in recovery of sparse vectors. In this paper, we use a class of DA functions which more closely approximate the delta function and show that this intuition is indeed the case. Fig.~\ref{fig:FigFunc} shows the above DA functions as well as $1 - \exp(-|x|/\sigma)$, one of the exploited DA functions in this paper in which $\sigma$ is a parameter to control the fitting to the delta function. Obviously, by choosing a small enough $\sigma$, it is possible to have the best fit to the delta function among the plotted DA functions. Putting this DA function in the sparse recovery problem, we are proposing to solve
\begin{equation} \label{FminExp}
\min_{\xb} \summ \Big(1 - \exp(-|x_i|/\sigma)\Big)  \quad \text{s.t.} \quad \| \Ab\xb - \bb\| \leq \epsilon,
\end{equation}
where $\epsilon$ is equal to $0$ in the noise-free case, to obtain a solution to \eqref{l0min} or \eqref{l0minNoisy}.

\subsection{Properties of DA Functions}
To establish theoretical analysis for the proposed algorithm and derive efficient optimization methods, we need to impose some assumptions on the DA functions, summarized in the following property. 

\newtheorem{Prop1}{Property}
\begin{Prop1} \label{Prop1}
Let  $f: \Rbb \rightarrow [-\infty,\infty)$ and define $\fsig(x) \triangleq f(x/\sigma)$ for any $\sigma > 0$. The function $f$ is said to possess Property \ref{Prop1}, if
\begin{enumerate}
\item[(a)] $f$ is real analytic on $(x_0,\infty)$ for some $x_0 < 0$,

\item[(b)] $\forall x \geq 0$, $f''(x) \geq -c$, where $c  > 0$ is some constant,

\item[(c)] $f$ is concave on $\Rbb$,

\item[(d)] $f(x) = 0 \Leftrightarrow x = 0$,


\item[(e)] $\lim_{x\rightarrow +\infty}f(x) = 1$.

\end{enumerate}
\end{Prop1}

It follows immediately from Property \ref{Prop1} that 
$\{\fsig(|x|)\}$ converges pointwise to $1-\delta(x)$ as $\sigma\rightarrow 0^{+}$; i.e.,
\begin{equation*}
\lim_{\sigma \to 0^{+}} \fsig(|x|)= \left\{
	\begin{IEEEeqnarraybox}[][c]{l?s}
    \IEEEstrut
	0 & if  $x = 0$\\
	1 & otherwise.
	\IEEEstrut
    \end{IEEEeqnarraybox}
    \right.
\end{equation*}

Besides the DA function $f(x) = 1-e^{-x}$ which is mainly used in this paper, there are other functions that satisfy conditions of Property \ref{Prop1} including
\begin{equation*}
f(x) = \left\{
    \begin{IEEEeqnarraybox}[][c]{l?s}
    \IEEEstrut
	\dfrac{x}{x+1} &  if $x \geq x_0$\\
	\,\!- \infty & otherwise
	\IEEEstrut
    \end{IEEEeqnarraybox}
    \right.
\end{equation*}
for some $-1 < x_0 < 0$.

\subsection{The Main Idea}
To obtain higher performance in recovery of sparse vectors, we propose to closely approximate the $\ell_0$ norm and then solve the consequent optimization problem. More precisely, let $f(\cdot)$ denote a function possessing Property \ref{Prop1}, then we define $\Fsig(|\xb|) = \summ \fsig(|x_i|) \approx \| \xb \|_0$ and use 
\begin{equation} \label{Fmin}
\min_{\xb} \Big( \Fsig(|\xb|) = \summ \fsig(|x_i|) \Big)  \quad \text{subject to} \quad \Ab\xb=\bb
\end{equation}
to find a solution to \eqref{l0min} and
\begin{equation} \label{FminNoisy}
\min_{\xb} \Fsig(|\xb|)  \quad \text{subject to}  \quad \| \Ab\xb - \bb\| \leq \epsilon
\end{equation}
to obtain a solution to \eqref{l0minNoisy}. 
Expectedly, these optimization problems are not convex, and any algorithm may get stuck in a local minimum.

Intuitively, when $\sigma$ is small and $\Fsig(|\xb|)$ approximates $\| \xb \|_0$ with a good accuracy, there are many local solutions which makes the task of optimizing \eqref{Fmin} and \eqref{FminNoisy} very hard. In contrast, if $\sigma$ is relatively large, while the accuracy of the approximation is not good, $\Fsig(|\xb|)$ has a smaller number of local minima. In line with this, in the asymptotic case, it will be shown that $\sigma \Fsig(|\xb|)$ becomes convex as $\sigma$ tends to infinity.

Following the same approach as in \cite{MohiBJ09,TrzaM09,MaleBAJ14}, a continuation scheme for solving \eqref{Fmin} and \eqref{FminNoisy} is utilized which helps in achieving the sparsest solution of $\Ab \xb = \bb$ and $\{ \xb~|~ \| \Ab\xb - \bb\| \leq \epsilon\}$ instead of finding a local minimum. Initially, optimization of \eqref{Fmin} or \eqref{FminNoisy} is started with a very large value of $\sigma$, and the solution is passed as an initial guess to the next iteration in which \eqref{Fmin} or \eqref{FminNoisy} is solved for a smaller value of $\sigma$. These iterations continue until reaching a desired accuracy.

To further decrease the chance of getting trapped in local minima, we constrain $\fsig(\cdot)$ to be continuous with respect to $\sigma$ in Property \ref{Prop1}. In this fashion, we expect that when $\sigma$'s at two consecutive iterations are close, the global minimizers 
for these two $\sigma$'s are in the vicinity of each other. Thus, starting from a convex optimization and gradually decreasing $\sigma$, it is more likely that a global solution 
will be found.

\section{Implementation of the SCSA algorithm} \label{sec:algoimp}

Contrary to \cite{TrzaM09} that tries to solve \eqref{Fmin} by converting it to an unconstrained problem and smoothing the DA function around the origin, we solve \eqref{Fmin} directly by employing a majorize-minimize (MM) technique \cite{HuntL04} without smoothing the DA function. This approach leads to iteratively solving a few weighted $\ell_1$ minimization (or linear) programs. Inspired by the iterative thresholding (IT) technique, we also propose two efficient methods for the noisy case of \eqref{FminNoisy} which are computationally attractive.

\subsection{Optimization for a Fixed $\sigma$ in the Noise-Free Case}

Although $\Fsig(|\xb|)$ is not a differentiable function, by restricting $\xb$ to be in the positive orthant, we can drop $|\cdot|$ from its argument, and make it differentiable. In other words, if we look for the sparsest solution which belongs to the positive orthant, then we can use $\fsig(x) = 1 - \exp(-x/\sigma)$ with the constraint $\xb \geq 0$ in \eqref{FminExp} and, in this way, make the cost function differentiable. Nevertheless, exploiting the same technique as in the conversion of \eqref{l1min} to a linear program \cite{Elad10}, we can overcome this restriction. To be precise, let 
$\yb = \begin{bmatrix} \xb_p^T ~ \xb_m^T \end{bmatrix}^T$ denote a column vector of length $2m$ in which $\xb_p = \max(\xb,\Zerb)$ and $\xb_m = - \min(\xb,\Zerb)$. The elements of $\yb$ are nonnegative, $\Fsig(\yb) = \Fsig(|\xb|)$, and the constraints $\Ab \xb = \bb$ are converted to $[\Ab,-\Ab]\yb=\bb$. By introducing this new vector, \eqref{Fmin} can be reformulated as
 \begin{equation} \label{FminPos}
\min_{\yb} \Fsig(\yb) \quad \text{subject to} \quad [\Ab,-\Ab]\yb=\bb,~\yb \geq \Zerb.
\end{equation}
The following theorem proves that, by solving \eqref{FminPos}, we are able to optimize program \eqref{Fmin}.

\newtheorem{Thm1}{Theorem}
\begin{Thm1} \label{EquivThm}
For any class of functions possessing Property \ref{Prop1}, programs \eqref{Fmin} and \eqref{FminPos} are equivalent.
\begin{proof}
The proof follows the same lines as in the proof of equivalence of $\ell_1$-minimization to a linear program \cite{Elad10}. Let 
$\ybs = \begin{bmatrix} \ub^T \vb^T \end{bmatrix}^T$, where $\ub,\vb \in \Rbb^m$, denote an optimal solution to \eqref{FminPos}. To show that \eqref{Fmin} and \eqref{FminPos} are equivalent, it is sufficient to show that the definition of $\yb$ in \eqref{FminPos} is not violated, or, mathematically speaking, the supports of $\ub$ and $\vb$ do not overlap. Assume, to the contrary, that they overlap at index $k$. Without loss of generality, further assume that $u_k > v_k$, then there is another solution $\widehat{\yb} = \begin{bmatrix} \widehat{\ub}^T \widehat{\vb}^T \end{bmatrix}^T$ with $\widehat{\ub} = \ub$ and $\widehat{\vb} = \vb$ except for $\widehat{u}_k = u_k - v_k$ and $\widehat{v}_k = 0$. This new solution 
 is feasible since $\widehat{\ub} \geq \Zerb$, $\widehat{\vb} \geq \Zerb$, and $[\Ab,-\Ab]\widehat{\yb}  = \bb$. However, $\Fsig(\widehat{\yb})$ is smaller than $\Fsig(\ybs)$ by $\fsig(u_k) + \fsig(v_k) - \fsig(u_k - v_k) > 0$ which contradicts the optimality of $\ybs$.
\end{proof}
\end{Thm1}

Since \eqref{FminPos} is a concave program, the MM technique can be easily used to find at least a local solution. First-order concavity condition for $\Fsig(\cdot)$ implies that
\begin{equation*}
\Fsig(\yb) \leq \Fsig(\ybt) + \langle \yb - \ybt, \nabla \Fsig(\ybt) \rangle = H_{\sigma}(\yb,\ybt)
\end{equation*}
for some $\ybt$ in the feasible set. To apply the MM technique, $H_{\sigma}(\cdot,\cdot)$ is selected as a surrogate function. Consequently, neglecting the fixed terms, to obtain a solution to \eqref{Fmin}, one needs to iteratively solve
\begin{equation} \label{FminPosLP}
\yb_{k+1} = \argmin_{\yb} \Big \{ \langle \nabla \Fsig(\yb_k),\yb \rangle ~\big|~ [\Ab,-\Ab]\yb=\bb,~\yb \geq \Zerb \Big\}
\end{equation}
for $k\geq 0$ until convergence. It can be verified that the program \eqref{FminPosLP} is equal to a weighted $\ell_1$ minimization
\begin{equation} \label{FminWL1}
\xb_{k+1} = \argmin_{\xb} \Big \{ \| \Wb \xb\|_1 ~\big|~ \Ab \xb=\bb \Big \},
\end{equation}
where $\Wb = \diag( \nabla \Fsig(\zb))$ and $\zb = |\xb_{k}|$.

The next proposition, which can be easily deduced from \cite[Theorem 3]{MaleBS14}, proves the convergence of the proposed MM based approach.

\newtheorem{Pro1}{Proposition}
\begin{Pro1} \label{LConvMM}
The sequence $\{\yb_k\}$ obtained from \eqref{FminPosLP} is convergent to a local minimum of \eqref{FminPos}, and $\Fsig(\yb_{k+1}) \leq \Fsig(\yb_{k})$ for all $k \geq 0$.
\end{Pro1}

\subsection{Optimization for a Fixed $\sigma$ in the Noisy Case}
A computationally attractive way to find a solution to an unconstrained optimization problem, in which the cost function is composed of the sum of a smooth and a nonsmooth convex function, is to use iterative thresholding or proximal algorithms; see \cite{CombP11} for a comprehensive discussion. This kind of problems is represented as
\begin{equation} \label{GenUn}
\min_{\xb} \lambda \rho(\xb) +  h(\xb),
\end{equation}
where $\rho(\xb)$ is convex but possibly nonsmooth, $h(\xb)$ is convex and differentiable with Lipschitz continuous gradient, and $\lambda$ is a regularization parameter. Particularly, program \eqref{l1minNoisy} can be converted to an equivalent unconstrained optimization problem, known also as the LASSO program \cite{BibkRA09},
\begin{equation} \label{l1minUn}
\min_{\xb} \lambda \|\xb\|_1  + \| \Ab\xb - \bb\|^2,
\end{equation}
where $\lambda > 0$ is a constant to regularize between solution sparsity and consistency to measurements. Accordingly, \eqref{l1minNoisy} can be solved by applying iterative thresholding  method on \eqref{l1minUn}. This special case of proximal methods is known as the iterative soft thresholding (IST) method. 
More generally, \eqref{l1minUn} can be written as
\begin{equation} \label{l1minUnGen}
\min_{\xb} \lambda \|\xb\|_1  + h(\xb),
\end{equation}
where $h(\xb)$ is as in \eqref{GenUn}. The IST method is aimed for finding a solution to \eqref{l1minUnGen} by iteratively solving \cite{BeckT09}
\begin{equation*}
\xb_{k+1} = \argmin_{\xb}\Big \{ \langle \xb - \xb_k, \nabla h(\xb_k) \rangle + \frac{1}{2\mu} \|\xb - \xb_k\|^2 + \lambda \|\xb\|_1 \Big \},
\end{equation*}
where $\mu$ is a step-size parameter. The above program admits a unique closed-form solution given by
\begin{equation} \label{l1minISTClosed}
\xb_{k+1} = \mathcal{S}_{\lambda \mu}(\xb_k - \mu \nabla h(\xb_k)),
\end{equation}
where $\mathcal{S}_{\alpha} : \Rbb \to \Rbb$ is the soft thresholding operator defined as $\mathcal{S}_{\alpha}(x) = \max(|x|-\alpha,0) \sign(x)$ for a scalar input and is applied component-wise to vectors.

To utilize the IT method for finding a solution to \eqref{FminNoisy}, first, program \eqref{FminNoisy} is formulated as an unconstrained optimization problem
\begin{equation} \label{FminUn}
\min_{\xb} \lambda_{\sigma} \Fsig(|\xb|) +  h(\xb),
\end{equation}
where $h(\xb)$ is as in \eqref{GenUn} and $\lambda_{\sigma}$ is a regularization parameter similar to \eqref{l1minUnGen} and, in general, may be a function of $\sigma$.

Similar to the convex case, 
\eqref{FminUn} can be optimized by iteratively solving
\begin{multline} \label{FminIST}
\xb_{k+1} = \\
\argmin_{\xb}\Big \{ \!\! \langle \xb - \xb_k, \nabla h(\xb_k) \rangle \! + \! \frac{1}{2\mu} \|\xb - \xb_k\|^2 \! + \! \lambda_{\sigma} \Fsig(|\xb|) \! \Big\}
\end{multline}
until convergence. Ignoring constant terms, it can be verified that the program \eqref{FminIST} is equal to
\begin{equation} \label{FminIST2}
\xb_{k+1} = \argmin_{\xb}\Big \{ \frac{1}{2\mu} \| \xb - (\xb_k - \mu \nabla h(\xb_k)) \|^2 + \lambda_{\sigma} \Fsig(|\xb|) \Big\}.
\end{equation}
For a general $\Fsig(\cdot)$, one should use iterative methods to solve \eqref{FminIST2}. However, remarkably, for  $\fsig(|x|) = 1 - \exp(-|x|/\sigma)$, we can find a closed-from solution using \emph{the Lambert W} function \cite{CorlGHJK96} which enables us to efficiently solve \eqref{FminIST2}. The details of derivation of the closed-form solution as well as the corresponding thresholding operator are given in Appendix \ref{appShr}. Using this thresholding operator, for any fixed $\sigma$, our IT based approach for solving \eqref{FminUn} simplifies to iteratively updating $\xb_{k+1}$ by
\begin{equation} \label{FminShr}
\xb_{k+1} = \mathcal{T}_{\lambda_{\sigma} \mu}^{(\sigma)}\big(\xb_k - \mu \nabla h(\xb_k)\big),
\end{equation}
where $\mathcal{T}_{\alpha}^{(\sigma)}$ is given in \eqref{ShrDef} in Appendix \ref{appShr}.

The next theorem whose proof is left for Appendix \ref{ISTConv} analyzes the convergence of the sequence generated by \eqref{FminIST} or \eqref{FminShr}.

\newtheorem{Thm2}[Thm1]{Theorem}
\begin{Thm1} \label{LConvIT}
Let $M > 0$ denote the smallest Lipschitz constant of $\nabla h$, and let $M'$ be the smallest value such that $\lambda_{\sigma} \nabla^{2} \Fsig(\xb) \succeq -M' \Ib, \forall \xb \geq \Zerb$.\footnote{For $h(\xb) = \| \Ab \xb - \bb\|^2$ and $\Fsig(\xb) = \sum_{i=1}^{m} 1 - e^{-x_i/\sigma}$, $M = 2 \lambda_{max}(\Ab^T \Ab)$ and $M' = \lambda_{\sigma} / \sigma^2$.} For any $\mu \in (0,\frac{1}{M + M'})$, the sequence $\{\xb_k\}$ generated by \eqref{FminIST} is convergent to a stationary point of \eqref{FminUn}.
\end{Thm1}

Although the IT technique has a low complexity, it lacks fast convergence rate \cite{BeckT09}. To speed up the rate of convergence, a Nesterov's like acceleration step is added in \cite{BeckT09} to the IST method which considerably increases the rate of convergence. While the complexity does not boost, improvement in the convergence rate is considerable. To accelerate our IT based approach, without proving the convergence, we also exploit a similar technique based on the FISTA algorithm \cite{BeckT09}. 
We call this instance of the SCSA algorithm fast iterative thresholding (FIT) based.

Now, let us remark on how the regularization parameter $\lambda_{\sigma}$ should be scaled with $\sigma$. For \eqref{FminUn} with $h(\xb) = \| \Ab \xb - \bb \|^2$, a necessary condition for optimality of a point $\xbs$ is that
\begin{equation} \label{FminUn_KKT}
\Zerb \in 2 \Ab^T \Ab \xbs - 2 \Ab^T \bb + \lambda_{\sigma} \partial \Fsig(|\xbs|)
\end{equation}
where $\partial \Fsig(|\xbs|)$ denotes the Clarke subdifferential \cite{Clar90} of $\Fsig(|\cdot|)$ at point $\xbs$. 
Particularly, in the scalar case, $\partial \fsig(|x|)$ for $\fsig(x) = 1 - e^{-x/\sigma}$ is given by
\begin{equation*}
\partial \fsig(|x|) = \left\{
\begin{IEEEeqnarraybox}[][c]{l?s}
\IEEEstrut
\Big\{ \frac{\sign(x)}{\sigma} e^{-\frac{|x|}{\sigma}} \Big\} & if $x \neq 0$ \\
\,\![-1/\sigma,1/\sigma] & if $x = 0$.
\IEEEstrut
\end{IEEEeqnarraybox}
\right.
\end{equation*}
Let $\tau,\xbs_{\tau},$ and $\Ab_{\tau}$ denote the support set of $\xbs$ and restriction of $\xbs$ and $\Ab$ to the entries and columns indexed by $\tau$, respectively. Condition \eqref{FminUn_KKT} implies that
\begin{IEEEeqnarray}{rCl} \label{NecCondImplicit}
& & \xbs_{\tau} = 
\Ab_{\tau}^\dagger \bb - \frac{\lambda_{\sigma}}{2\sigma} (\Ab_{\tau}^T \Ab_{\tau})^{-1} \sign(\xbs_{\tau}) \odot e^{-\frac{|\xbs_{\tau}|}{\sigma}},\\
& & |\ab_i^T (\Ab_{\tau} \xbs_{\tau} - \bb) | \leq \frac{\lambda_{\sigma}}{2\sigma},\quad i \not\in  \tau,\label{NecCondImplicit2}
\end{IEEEeqnarray}
where $\ab_i$ denotes the $i$th column of $\Ab$. The second term in the right-hand side of \eqref{NecCondImplicit} disappears when $\sigma \to 0$, since the exponential terms decay faster than $\sigma$. Therefore, if $\tau$ coincides with the support set of the true solution, \eqref{NecCondImplicit} shows that $\xbs_{\tau}$ tends to the oracle solution, which is obtained by knowing the support set of the true solution. However, when $\sigma$ is decreased to smaller values, inequality \eqref{NecCondImplicit2} will be satisfied easier, and \eqref{FminUn} will give solutions with smaller sparsity levels. In fact, $|\ab_i^T (\Ab_{\tau} \xbs_{\tau} - \bb) |$ measures how close are the residual $\Ab_{\tau} \xbs_{\tau} - \bb$ and the $i$th column of $A$; hence, the larger the threshold, the larger the number of indices of columns of $\Ab$ to be excluded from $\tau$. Moreover, the threshold $\lambda_{\sigma}/\sigma$ should naturally be independent of $\sigma$ but dependent on the noise level. Thus, to have this threshold independent of $\sigma$, $\lambda_{\sigma}$ is scaled linearly with $\sigma$; i.e., $\lambda_{\sigma} = \lambda \sigma$ for some constant $\lambda$.

\begin{figure}[tb]
\centering
\includegraphics[width=0.37\textwidth]{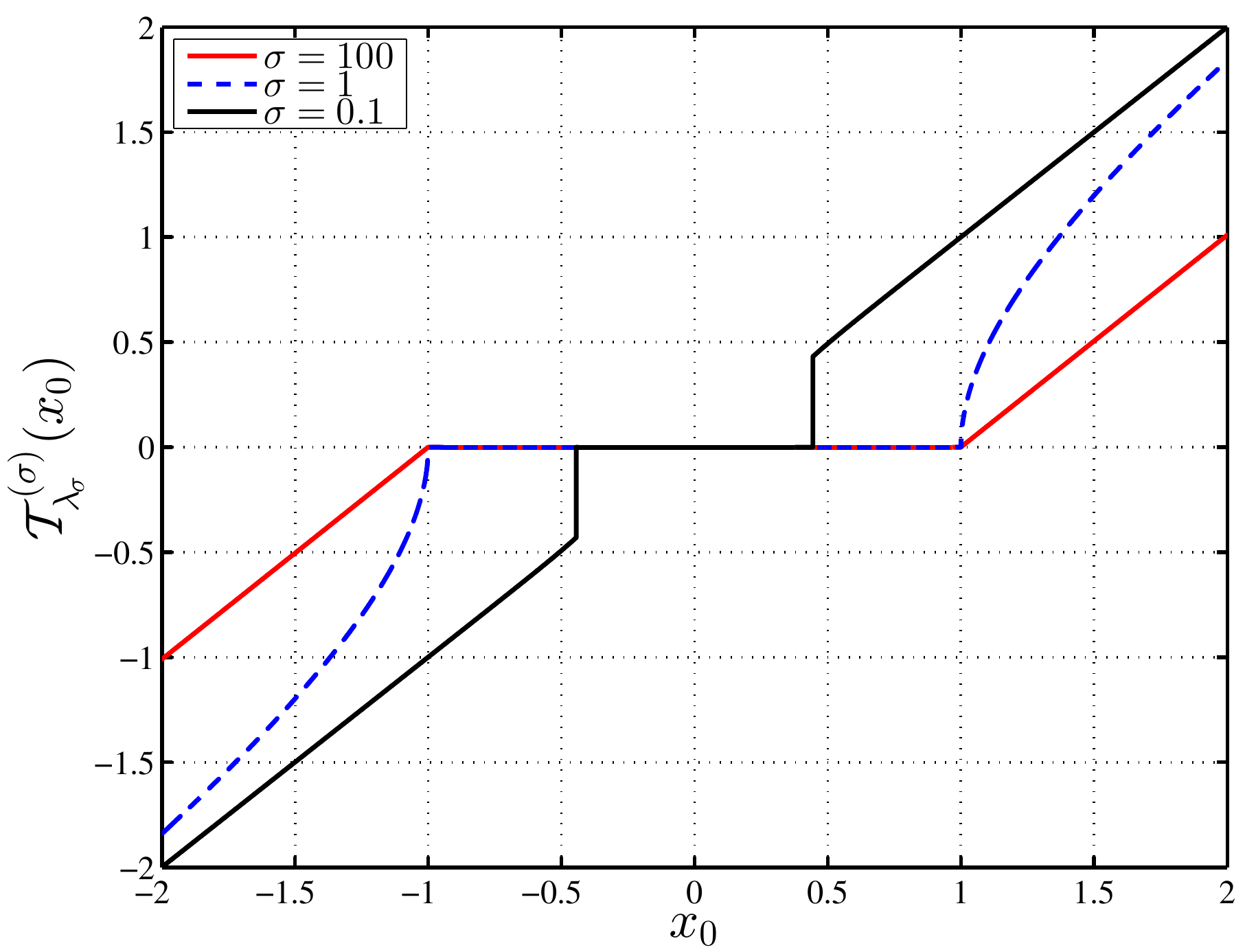}
\vspace{-0.1cm}
\caption{$\mathcal{T}_{\lambda_{\sigma}}^{(\sigma)}$ is plotted for $\sigma = 100,1, \text{ and } 0.1$ with $\lambda = 1$ and $\lambda_{\sigma} = \lambda \sigma$. For the large $\sigma$, $\mathcal{T}_{\lambda_{\sigma}}^{(\sigma)}$ is very close to the soft thresholing operator $\mathcal{S}_{\lambda}$ with $\lambda = 1$, whereas it is very similar to the hard thresholding operator defined in \eqref{HardThing} with $\lambda = 0.1$.} \label{fig:HardSoft}
\end{figure}

A closer look at the thresholding operator introduced in \eqref{ShrDef} reveals an interesting interpolation property. Formally, in Propositions \ref{InitThm} and \ref{Convl0Pro}, we will show how \eqref{Fmin} asymptotically converges to $\ell_0$ and $\ell_1$ minimization when $\sigma$ goes to $0$ or $\infty$. The thresholding operator $\mathcal{T}_{\lambda_{\sigma}}^{(\sigma)} (\cdot)$, however, also more simply illustrates this asymptotical behavior for $1 - \exp(-|x|/\sigma)$. Figure \ref{fig:HardSoft} displays $\mathcal{T}_{\lambda_{\sigma}}^{(\sigma)} (\cdot)$ for $\sigma = 100,1$, and $0.1$ when $\lambda_{\sigma} = \lambda \sigma$ and $\lambda$ is fixed to 1. In this plot, when $\sigma$ is relatively large, $\mathcal{T}_{\lambda_{\sigma}}^{(\sigma)} (\cdot)$ is very close to the soft thresholding operator, whereas, for a small $\sigma$, it is very similar to the hard thresholding operator \cite{BlumD09}, which according to the formulation used in this paper is defined as
\begin{equation} \label{HardThing}
\mathcal{H}_{\lambda}(x) = |x| \mathbbm{1}_{|x| > \sqrt{2\lambda}},
\end{equation}
where $\mathbbm{1}$ denotes the indicator function. This shows that when $\sigma$ is swept from very large to smaller values, the thresholding operator gradually converts from the soft thresholding operator to the hard thresholding operator, making an interpolation between $\ell_1$ and $\ell_0$ minimization.

{\color{\ROne_Col}Finally, it is worth comparing our above proposed approach to a few other available methods. In \cite{Zhan10} and \cite{Zhan13}, a multi-stage convex relaxation method for solving \eqref{FminUn} has been proposed, where the nonconvex function $F(\cdot)$ is more general than those defined by Property \ref{Prop1}.\footnote{{\color{\ROne_Col}In the formulation of \cite{Zhan10}, $F(\cdot)$ also does not depend on any scaling parameter like $\sigma$.}} When $F(\cdot)$ is aimed for sparsity regularization, the proposed optimization method involves solving a number of weighted versions of \eqref{l1minUn}. Namely, one needs to iteratively solve
\begin{equation*}
\xb_{k+1} = \argmin_{\xb} \Big \{ \lambda \|\Wb_{k} \xb\|_1  + \| \Ab \xb - \bb \|^2 \Big \},
\end{equation*}
where $h(\xb)$ is assumed to be equal to $\| \Ab \xb - \bb \|^2$ and $\Wb_{k}$ is the weighting matrix that depends on $\xb_k$ (the previous solution) and the function $F(\cdot)$ \cite{Zhan10,Zhan13}. 
In contrast to this approach and instead of solving a sequence of optimization problems to obtain a solution to \eqref{FminUn}, our proposed method directly solves \eqref{FminUn} which in turn seems to be more computationally efficient. Moreover, in \cite{WangLZ14}, a regularization path-following algorithm for solving \eqref{FminUn} is introduced which is based on the iterative thresholding approach. Along with this algorithm, some strong theoretical guarantees are provided for a special class of nonconvex regularizations. This algorithm as well as its theoretical analyses, however, cannot be applied to the DA functions considered herein since, for instance, $\fsig(|x|) = 1 - \exp(-|x|/\sigma)$ does not satisfy regularity condition (a) in \cite{WangLZ14}.}

\subsection{Initialization}
The following proposition proves that when $\sigma \to \infty$, a scaled version of $\Fsig(|\xb|)$ becomes equal to the $\ell_1$ norm. Consequently, 
the MM based instance of the proposed algorithm is initialized with a minimum $\ell_1$-norm solution of $\Ab \xb = \bb$, and the IT and FIT based instances are initialized with the solution obtained by the FISTA algorithm. The proof easily follows from \cite[Theorem 1]{MaleBS14}.
\newtheorem{Pro2}[Pro1]{Proposition}
\begin{Pro2} \label{InitThm}
For any class of functions $\{\fsig\}$ possessing Property \ref{Prop1},
\begin{equation*}
\lim_{\sigma \to \infty} \frac{\sigma}{\gamma} \Fsig(|\xb|) =  \| \xb \|_1,
\end{equation*}
where $\gamma = f'(0) \neq 0$.
\end{Pro2}


\makeatletter
\renewcommand{\ALG@name}{Subalgorithm}
\makeatother
\begin{algorithm}[t] \small
\caption{Opt\_FIT: Optimization of \eqref{FminUn} for a fixed $\sigma$ using the FISTA like acceleration.}\label{alg:SCSA_FIT}
\mbox{Input: $\Ab, \bb, \lambda, \mu, \sigma, \xb_0,$ \texttt{tol}} \newline
\mbox{Initialization:}
\begin{algorithmic}[1]
\STATE $\yb_1 = \xb_0,~t_1 = 1,~k=0,~h(\xb) = \| \Ab \xb - \bb \|^2$.
\end{algorithmic}  %
\mbox{Body:}%
\begin{algorithmic}[1]%
\WHILE{$d > \texttt{tol}$}
\STATE $k = k + 1$.
\STATE $\xb_k = \mathcal{T}_{\lambda \mu}^{(\sigma)}(\yb_{k} - \mu \nabla h(\yb_k)).$
\STATE $t_{k+1} =  (1 + \sqrt{1 + 4t_{k}}) / 2.$
\STATE $\yb_{k+1} =  \xb_{k}+ ( t_{k} - 1) (\xb_{k} -\xb _{k-1}) / t_{k+1}$.
\STATE $d = \| \xb_{k}  - \xb_{k-1} \| / \| \xb_{k-1} \|$.
\ENDWHILE
\end{algorithmic}
\mbox{Output: $\xb_{k}$}
\end{algorithm}


\addtocounter{algorithm}{-1}
\makeatletter
\renewcommand{\ALG@name}{Algorithm}
\makeatother
\begin{algorithm}[t] \small
\caption{The SCSA algorithm}\label{alg:SCSA}
\mbox{Input: $\Ab, \bb, \lambda$ (Only for IT and FIT based instances)} \newline
\mbox{Initialization:}
\begin{algorithmic}[1]
\STATE $c \in (0,0.5)$: decreasing factor for $\sigma$.
\STATE $\epsilon_1$, $\epsilon_2 > 0$: stopping thresholds for main and internal loops.
\STATE $h(\xb)= \| \Ab \xb - \bb \|^2, \fsig(x) = 1 - e^{-|x|/\sigma}$.
\vspace{0.3em}
\STATE $\left\{
\!\!
\begin{array}{ll}
	\text{LP based:} & \xb_0 = \argmin_{\xb} \big\{ \|\xb\|_1 | \Ab \xb = \bb \big\}.\\
	\text{IT or FIT based:} & \xb_0 = \argmin_{\xb} \big\{\lambda \| \xb\|_1 + \| \Ab \xb - \bb \|^2 \big\}.
	\end{array}
\right.$
\STATE $\sigma_0 = 8 \max(|\xb_0|)$.
\end{algorithmic}
\mbox{Body:}
\begin{algorithmic}[1]
\STATE $i = 0, \sigma = \sigma_0.$
\WHILE{$d_1 > \epsilon_1$}
\STATE $\xbh_0 = \xb_i, j = 0, i=i+1,\lambda_{\sigma} = \lambda \sigma.$
    \STATE \textbf{while} $d_2 > \epsilon_2$ \textbf{do}
    \STATE \hspace{\algorithmicindent} $j = j + 1.$
    \STATE \hspace{\algorithmicindent} $\Wb = \diag( \nabla \Fsig(|\xbh_{j-1}|))$.
    \vspace{0.3em}
    \STATE \hspace{\algorithmicindent}  \!\!\! $\left\{
                                        \!\!\!
                                        \begin{array}{ll}
	                                    \text{LP based:} & \xbh_j = \argmin_{\xb} \big \{ \| \Wb \xb\|_1 ~\big|~ \Ab \xb=\bb \big \}.\\
	                                    \text{IT based:} & \xbh_{j} = \mathcal{T}_{\lambda_{\sigma} \mu}^{(\sigma)}\big(\xbh_{j-1} - \mu \nabla h(\xbh_{j-1})\big).\\
                                        \text{FIT based:} & \xbh_j = \text{Opt\_FIT}(\Ab, \bb, \lambda_{\sigma}, \mu, \sigma, \xbh_{j-1}, \epsilon_2).\\
	                                    \end{array}
\right.$
    \STATE \hspace{\algorithmicindent} $d_2 = \| \xbh_{j}  - \xbh_{j-1} \| / \| \xbh_{j-1} \|.$

    \STATE \textbf{end while}
\STATE $\xb_{i} = \xbh_{j}.$
\STATE $d_1 = \| \xb_{i}  - \xb_{i-1} \| / \| \xb_{i-1} \|.$
\STATE $\sigma=c \sigma.$
\ENDWHILE
\end{algorithmic}
\mbox{Output: $\xb_i$}
\end{algorithm}

\subsection{The Final Algorithm}
Putting all the above steps together, the final algorithm, summarized in Algorithm \ref{alg:SCSA}, is obtained by exploiting $\fsig(x) =  1 - e^{-x/\sigma}$ and $h(\xb) = \| \Ab \xb - \bb\|^2$. According to the method used in optimizing \eqref{Fmin} or \eqref{FminUn} for a fixed $\sigma$, three instances of the SCSA algorithm are summarized as
\begin{itemize}
  \item SCSA-LP (based on linear programming),
  \item SCSA-IT (based on IT method),
  \item SCSA-FIT (based on FIT method).
\end{itemize}
To completely characterize the implementation of this algorithm, the following remarks are in order.


\emph{Remark 1.} Parameter $\sigma$ is decayed by a multiplicative factor $c$ which should be chosen in the interval $(0,0.5)$. We will discuss how to properly choose it in Section \ref{sec:NumExp} with more details. Furthermore, following the same reasoning as in \cite{MaleBS14}, $\sigma_0$ is set to $8 \max(|\xb_0|)$ because this $\sigma_0$ virtually acts as if $\sigma$ tends to $\infty$.

\emph{Remark 2.} $d_1 = \| \xb_{i}  - \xb_{i-1} \| / \| \xb_{i-1} \|$ and $d_2 = \| \xbh_{j}  - \xbh_{j-1} \| / \| \xbh_{j-1} \|$ measure relative distances between the solution of successive iterations of the external and internal loops, respectively, and are used to stop execution of these loops. In the proposed continuation approach, it is not necessary to run the internal loop until convergence, and it is just needed to get close to the minimizer for the current value of $\sigma$. Consequently, $\epsilon_1$ is usually set a few orders of magnitude smaller than $\epsilon_2$. For the noise-free case, $\epsilon_1 = 10^{-3}$ and $\epsilon_2 = 10^{-2}$ are suggested. In the noisy case, since the IT and FIT based approaches are relatively slow and the difference between two consecutive solutions is not large, $\epsilon_1$ and $\epsilon_2$ should be chosen smaller. Moreover, as SCSA-IT generally has a slower convergence rate, $\epsilon_2$ (the threshold for the internal loop) for this instance of SCSA should be smaller than that of SCSA-FIT. This choice, as will be shown in numerical simulations, leads to a similar performance in terms of reconstruction accuracy. $\epsilon_1$ and $\epsilon_2$ are also functions of the regularization parameter $\lambda$. Putting altogether, we numerically found that a good choice for SCSA-IT is $\epsilon_1 = \min(10^{-4},10^{-3}\lambda)$ and $\epsilon_2 = \min(10^{-4},10^{-3}\lambda)$ and for SCSA-FIT is $\epsilon_1 = \min(10^{-4},10^{-3}\lambda)$ and $\epsilon_2 = \min(10^{-3},10^{-2}\lambda)$. 

\emph{Remark 3.} Proposition \ref{InitThm} can be strengthened to
\begin{equation*}
\lim_{\sigma \to \infty} \argmin_{\xb} \big\{\Fsig(|\xb|)|\Ab \xb = \bb\big\} = \argmin_{\xb} \big\{\|\xb\|_1|\Ab \xb = \bb\big\}
\end{equation*}
provided that the above $\ell_1$ minimization has a unique solution. Nevertheless, since not a strictly convex program, it may occur that $\ell_1$ minimization does not admit a unique solution. Let $S_0$ and $S_1$ denote the solution sets of \eqref{l0min} and \eqref{l1min}, respectively. An interesting problem is to characterize the conditions under which $S_0$ (which we assume to be singleton) is a subset of $S_1$. Given these conditions, one can hope to devise a suitable optimization algorithm which is theoretically guaranteed to start from a point in $S_1$ and end up in the unique solution of \eqref{l0min}. In this fashion, the guaranteed recovery bounds for $\ell_1$ minimization can be improved.

\emph{Remark 4.} As discussed earlier, $\lambda_{\sigma}$ is set to $\lambda \sigma$ for some $\lambda > 0$. This choice can be also justified as follows. Since SCSA in the noisy case is initialized with a solution to \eqref{l1minUn} and $\lim_{\sigma \to \infty} \sigma \Fsig(|\xb|) = \| \xb \|_1$, 
it is natural to set $\lambda_{\sigma} = \lambda \sigma$ to have the same cost function in \eqref{l1minUn} and \eqref{FminUn} for $\sigma \to \infty$.

\section{Theoretical Analysis} \label{sec:PerAn}
{\color{\RTwo_Col}A thorough performance analysis of the SCSA algorithm considering all of its steps seems to be very hard and cannot be embedded in this paper. In this section, however, by analyzing programs \eqref{Fmin} and \eqref{FminNoisy} for any $\sigma > 0$ and/or $\sigma$ tending to $0^{+}$, we provide simplified analyses which will give the reader a theoretical insight about how the main idea works. }
These analyses are simply extracted from the results in \cite{GribN07,Zhan10RIP,MaleBS14,LiuJG13} and are based on the null-space \cite{GribN07}, restricted isometry \cite{Cand08}, and spherical section properties \cite{Zhan08,KashT07} of the sensing matrix. We recall or modify them in order to be able to study the performance of the proposed algorithm.

\emph{Null-space based recovery conditions:} It can be verified that Property \ref{Prop1} implies the `sparseness measure' definition in \cite{GribN07}. Consequently, based on Theorems 2 and 3 and Lemma 4 of \cite{GribN07}, a necessary and sufficient condition for exact recovery of sparse vectors via \eqref{Fmin} is as follows. Let us define
\begin{equation*}
\theta_{\fsig}(s,\Ab) \triangleq \sup_{\hb \in \NullDefmZ} \frac{\sum_{i=1}^{s} \fsig(h_i^{\downarrow})}{\sum_{i=1}^{m} \fsig(h_i)},
\end{equation*}
where $h_i^{\downarrow}$ denotes the $i$th largest (in magnitude) component of $\hb$. $\theta_{\fsig}(s,\Ab) < 1/2$ is a necessary and sufficient condition for exact recovery of all vectors with sparsity at most $s$. These conditions are weaker than those corresponding to $\ell_1$ minimization and lead to the following proposition which is a special case of \cite[Proposition 5]{GribN07}.

\newtheorem{Pro3}[Pro1]{Proposition}
\begin{Pro3} \label{Sup}
Let $m_{\fsig}^*(\Ab)$, $m_{\ell_0}^*(\Ab)$, and $m_{\ell_1}^*(\Ab)$ denote the maximum sparsity such that all vectors $\xb$ with $\|\xb\|_0 \leq m_{\fsig}^*(\Ab)$, $\|\xb\|_0 \leq m_{\ell_0}^*(\Ab)$, and $\|\xb\|_0 \leq m_{\ell_1}^*(\Ab)$ can be uniquely recovered by \eqref{Fmin}, \eqref{l0min}, and \eqref{l1min}, respectively. For any $\fsig(\cdot)$ possessing Property \ref{Prop1},
\begin{equation*}
m_{\ell_1}^*(\Ab) \leq m_{\fsig}^*(\Ab) \leq m_{\ell_0}^*(\Ab).
\end{equation*}
\end{Pro3}

The so-called robust recovery condition is satisfied if all vectors with sparsity at most $s$ can be recovered from \eqref{FminNoisy} with an error proportional to $\epsilon \geq \| \wb \|$ \cite{LiuJG13}. In general, extension of the above necessary and sufficient conditions to robust recovery of sparse vectors from noisy measurement is not easy. However, \cite{LiuJG13} proves that, under some mild assumptions, the sets of sensing matrices satisfying the exact and robust recovery conditions differ by a set of measure zero. In other words, $\theta_{\fsig}(s,\Ab) < 1/2$ also guarantees the accurate recovery of sparse vectors via \eqref{FminNoisy} for most sensing matrices.

\emph{Restricted isometry property based conditions:} Let $\delta_{2s}$ and $\delta_{3s}$ denote the restricted isometry constants of orders $2s$ and $3s$ defined in \cite{Cand08}. For a general concave function $\fsig(\cdot)$ (including those satisfying Property \ref{Prop1}), \cite{Zhan10RIP} shows that $\delta_{2s} < 1/2$ and $\delta_{3s} < 2/3$ are sufficient for accurate recovery of a sparse vector with the $\ell_0$-norm not greater than $s$ via \eqref{FminNoisy}.

\emph{Spherical section property based conditions:} While the above recovery conditions do not provide strict superiority to $\ell_1$ minimization, the following proposition shows that, in the noise-free case, one can obtain a solution arbitrarily close to the unique solution of $\ell_0$ minimization by properly choosing $\sigma$. Indeed, as long as $\ell_0$ minimization admits a unique solution, it is possible to recover it by \eqref{Fmin} letting $\sigma \to 0$. However, it is obvious that, for sufficiently sparse vectors, $\theta_{\fsig}(s,\Ab) < 1/2$ guarantees exact recovery for any $\sigma > 0$. To state the result, first, the definition of the spherical section property is recalled.

\newtheorem{Def1}{Definition}
\begin{Def1}[Spherical Section Property \cite{Zhan08,KashT07}]
The sensing matrix $\Ab$ possesses the $\Delta$-spherical section property if, for all $\wb \in \NullDefmZ$, $\|\wb\|_1^2/\|\wb\|^2 \geq \Delta(\Ab)$. In other words, the spherical section constant of the matrix $\Ab$ is defined as
\begin{equation*}
\Delta(\Ab) \triangleq \min_{\wb \in \nullS(\Ab) \setminus \{\Zerb\}} \frac{\|\wb\|_1^2}{\|\wb\|^2}.
\end{equation*}
\end{Def1}

It is known that many randomly generated sensing matrices possess the SSP with high probability \cite{Zhan08}. The proof of the following proposition easily follows from Proposition 4 of \cite{MaleBS14} by restricting the matrices to be diagonal.

\newtheorem{Pro4}[Pro1]{Proposition}
\begin{Pro4}\label{Convl0Pro}
Assume that $\Ab \in \Rbb^{n \times m}$ has the $\Delta$-spherical property, and consider a class of functions $\{\fsig\}$ possessing Property \ref{Prop1}. Let $\xbsi$ denote a solution to \eqref{Fmin}, and let $\xb_0$ denote the unique solution to \eqref{l0min}. Then
\begin{equation*}
\|\xbsi - \xb_0\| \leq \frac{n \alpha_{\sigma}}{\sqrt{\Delta}-\sqrt{\UDelmOne}},
\end{equation*}
where $\alpha_{\sigma} = \big| \fsig^{-1}(1-\frac{1}{n}) \big|$. The above inequality also leads to
\begin{equation*}
\lim_{\sigma \to 0^{+}} \xbsi = \xb_{0}.
\end{equation*}
\end{Pro4}

\section{Numerical Experiments} \label{sec:NumExp}
To assess the effectiveness of the SCSA algorithm in recovering sparse vectors, a number of numerical experiments are performed. Initially, the effect of parameter $c$ is examined, and a suitable choice for this parameter is proposed. Next, the performance of SCSA in 
noiseless and noisy settings is evaluated and compared to some of the state-of-the-art algorithms. The general experimental setups as well as specific settings for the noise-free and noisy cases are described in the following subsection.

\subsection{Experimental Setups}
We use randomly generated sparse vectors and sensing matrices in all numerical experiments. More specifically, following common practice, each entry of the sensing matrix $\Ab$ is generated independently from the zero-mean, unit-variance Gaussian distribution $\ensuremath{N(0,1)}$, 
and the columns of $\Ab$ are normalized to have unit $\ell_2$-norm. 
To construct a sparse vector $\xbt$ with $\| \xbt \|_0 = s$, first, the location of nonzero components is sampled uniformly at random among all possible subsets of $\{ 1,\cdots,m\}$ with cardinality $s$. Then the values of nonzero components are drawn independently from either $\ensuremath{N(0,1)}$ or the Rademacher distribution of $\{\pm1\}$ with equal probability.
Moreover, when measurements are noisy, the noise vector $\wb$ is always drawn from $\ensuremath{N(0,\sigma^2_{w}\Ib})$. Finally, the vector of measurements $\bb$ is equal to  $\Ab \xbt + \wb$, where $\wb = \Zerb$ when the noise-free case is under consideration.

Let $\xbh$ denote the output of one of the algorithms used in the numerical experiments to recover the sparse vector $\xbt$ from either noisy or noiseless measurements. 
We use the following {\color{\ROne_Col}four} quantities to measure and compare reconstruction accuracy in different experiments. 
\begin{itemize}
  \item Reconstruction SNR in dB:\\ $\SNR \triangleq 20 \log_{10}(\| \xbt \| / \| \xbt - \xbh \|)$ which will be used in Experiment 1 and, implicitly, in Experiment 2.
  \item Median reconstruction SNR:\\ $\MSNR \triangleq 10 \log_{10} (\| \xbt \|^2 / \median(\| \xbt - \xbh \|^2))$ where $\median(\| \xbt - \xbh \|^2)$ denotes the median of $\| \xbt - \xbh \|^2$ over all the Monte-Carlo simulations. $\MSNR$ is used in Experiments 3 and 4.
  \item {\color{\ROne_Col}Support recovery rate (SRR):\\ Let $\widetilde{\tau}$ and $\widehat{\tau}$ denote the support set of $\xbt$ and the set of indices of the $s$ largest (in magnitude) components of $\xbh$, respectively. SRR is defined as the number of realizations in which $\widetilde{\tau} = \widehat{\tau}$ normalized by the total number of Monte-Carlo simulations. SRR is used in Experiment 4.}
  \item Mean-squared error (MSE) which is the sample mean of $\| \xbt - \xbh \|^2$ and will be used in Experiment 5.

\end{itemize}
Besides the accuracy, execution time, as a rough measure of the computational complexity, is used to compare algorithms. All simulations are performed in MATLAB 8 environment using an Intel Core i7{\color{\MinEdt}-4600U, 2.1} GHz processor with 8 GB of RAM, under Microsoft Windows 7 operating system.

\subsubsection{noise-free case}
An algorithm is declared to be successful in recovering the solution, if 
$\SNR \geq 60\text{ dB}$. Consequently, to compare the performance of different algorithms, we use the success rate defined as the number of times an algorithm successfully recovers the solution divided by the total number of trials. To solve \eqref{l1min} and \eqref{FminWL1}, where the latter is also used in the implementation of some of the algorithms in the comparison with an algorithm-dependent weighting matrix, we use the $\ell_1$-magic \cite{CandR05}.

\subsubsection{noisy case}
We assume that the noise variance, $\sigma_w^2$, is known; thus, it is possible to use the following formula \cite{BellCW11,BibkRA09}
\begin{equation} \label{LASSOlam}
\lambda = 2 c_{r} \sigma_w \Phi^{-1}(1 - \frac{\alpha_{r}}{2m})
\end{equation}
to choose the regularization parameter for the LASSO estimator \eqref{l1minUn}. With the above choice, in which $c_r > 1$ is some constant and $\Phi$ is the cumulative density function of $\ensuremath{N(0,1)}$, the LASSO estimator achieves the so-called `near-oracle' performance with probability at least $1-\alpha_r$ \cite{BellCW11}. \footnote{As will be explained later in Subsection \ref{sec:NoisyExp}, with this choice of $\alpha_r$ which is the same as in \cite{BenEE10}, we are evaluating the typical performance of the LASSO, SCSA-IT, and SCSA-FIT algorithms.} In our numerical experiments, $c_{r}$ and $\alpha_{r}$ are set to 1.05 and 0.5, respectively. The above regularization parameter is used for IST and FISTA based implementations of the LASSO as well as IT and FIT instances of the SCSA algorithm.

Finally, it should be mentioned that, to have more stable plots without large fluctuations, in the noisy case, we normalize the $\ell_2$ norm of each $s$-sparse vector to $\sqrt{s}$.


\subsection{Effect of Parameter $c$}
\emph{Experiment 1.} To increase the accuracy of approximating the $\ell_0$ norm in SCSA, $\sigma$ is decreased according to the rule $\sigma_i = c \sigma_{i-1}, i \geq 1$. Intuitively, a small value for $c$ corresponds to fast decay of $\sigma$ and increase in the risk of getting trapped in local solutions. In contrast, a relatively large value of $c$ leads to smooth changes in $\Fsig(\cdot)$ where it is more likely to end up in the global solution. 
In the simplest case, where the sparsity of the solution is small enough and measurements are noise-free, 
the sufficient condition stated in Section \ref{sec:PerAn} implies that \eqref{l1min} and \eqref{Fmin} have the same unique solution. As SCSA is initialized with the minimum $\ell_1$-norm solution and Proposition \ref{LConvMM} proves that, for any $\sigma$, $\Fsig(\cdot)$, in the internal loop, is nonincreasing, the SCSA algorithm converges after 2 iterations independent of $c$. Moreover, in the noisy case, from the theoretical analysis presented in Section \ref{sec:PerAn} and the proof of Theorem \ref{LConvIT} which shows that $\lambda_{\sigma} \Fsig(|\xb_i|) +  h(\xb_i)$ is not increasing in $i$, we expect a similar behavior. On the other hand, when the sparsity level increases, to decrease the risk of getting trapped in local minima, a larger $c$ should be selected.

\begin{figure}[tb]
\centering
\includegraphics[width=0.49\textwidth]{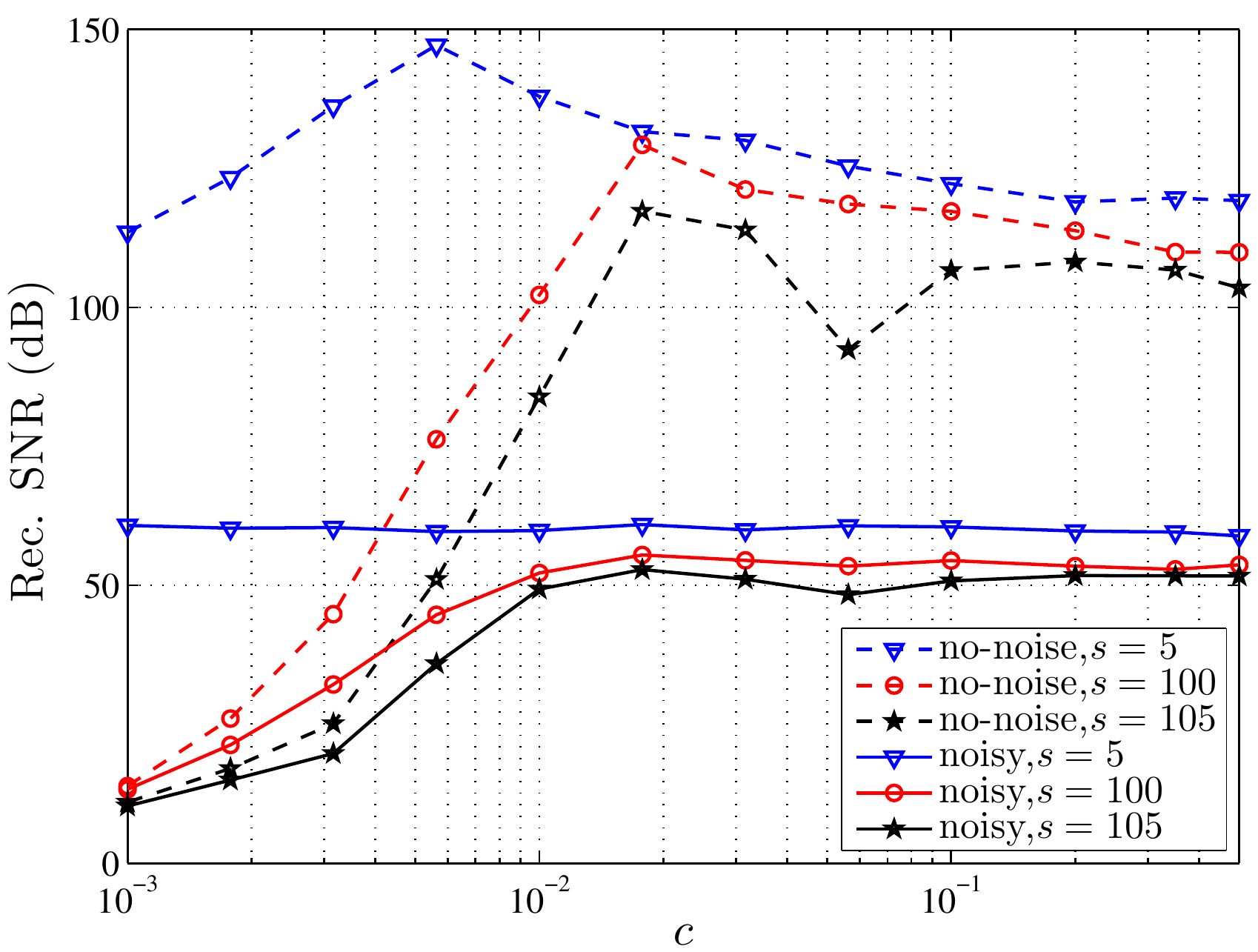}
\vspace{-0.6cm}
\caption{Averaged $\SNR$ of the SCSA algorithm in solving noiseless and noisy problems are plotted versus $c$ for 3 different sparsity levels. The sparse vector is of length 400, and the sensing matrix is $200 \times 400$. To have an accurate estimate of the $\SNR$, in each problem, 100 Monte-Carlo simulations are run, and results are averaged.} \label{fig:cEff}
\end{figure}

To see the above intuition, the effect of parameter $c$ in the reconstruction SNR in noisy and noiseless cases is numerically experimented. The dimensions of the sensing matrix are fixed to $200 \times 400$, $\sigma_w$ is equal to $10^{-3}$, and the sparse vectors are Gaussian distributed. The experiment is repeated for 3 different cardinalities ($s=5,100,105$), and $\SNR$ is averaged over 100 trials. Fig.~\ref{fig:cEff} shows the averaged $\SNR$'s as a function of $c$. As predicted, when $s$ is small, the $\SNR$ is always high. On the other hand, for high sparsity, after passing a critical value, $\SNR$ remains almost unchanged. Based on this observation, in the remaining experiments, we conservatively set $c$ to 0.1.

\subsection{Noise-Free Recovery}
\emph{Experiment 2.} In this experiment, the performance of the SCSA algorithm in the noise-free setting is compared to $\ell_1$ minimization, the SL0 algorithm \cite{MohiBJ09}, $\ell_p$ quasi-norm minimization \cite{FoucL09,Char07}, and the reweighted $\ell_1$ minimization \cite{CandWB08} in terms of success rate and execution time. {\color{\MinEdt}The following implementation method and parameters are used for each algorithm.
\begin{itemize}
  \item When implementing the SCSA algorithm, $\epsilon_1$ and $\epsilon_2$ are set to $10^{-3}$ and $10^{-2}$, respectively, and $c$ is set to 0.1.
  \item For SL0 (MATLAB code: http://ee.sharif.edu/$_{\widetilde{~}}$SLzero/), the following parameters are used: \texttt{sigma\_min=10\textsuperscript{-4}}, \texttt{c=0.8}, \texttt{mu=2}, and \texttt{L=8}. As suggested in \cite{MohiBJ09}, these parameters result in a much better success rate than that provided by the default values.
  \item $\ell_p$ quasi-norm minimization is implemented based on the iteratively reweighted $\ell_1$ minimization approach in \cite{ChartY08} with $p = 0.5$.
  \item For the reweighted $\ell_1$ minimization \cite{CandWB08}, we use $\epsilon = 0.1$ which attains the best performance \cite{CandWB08}.
\end{itemize}
}

\begin{figure}[tb]
        \centering
        \subfigure{%
                \includegraphics[width=0.49\textwidth]{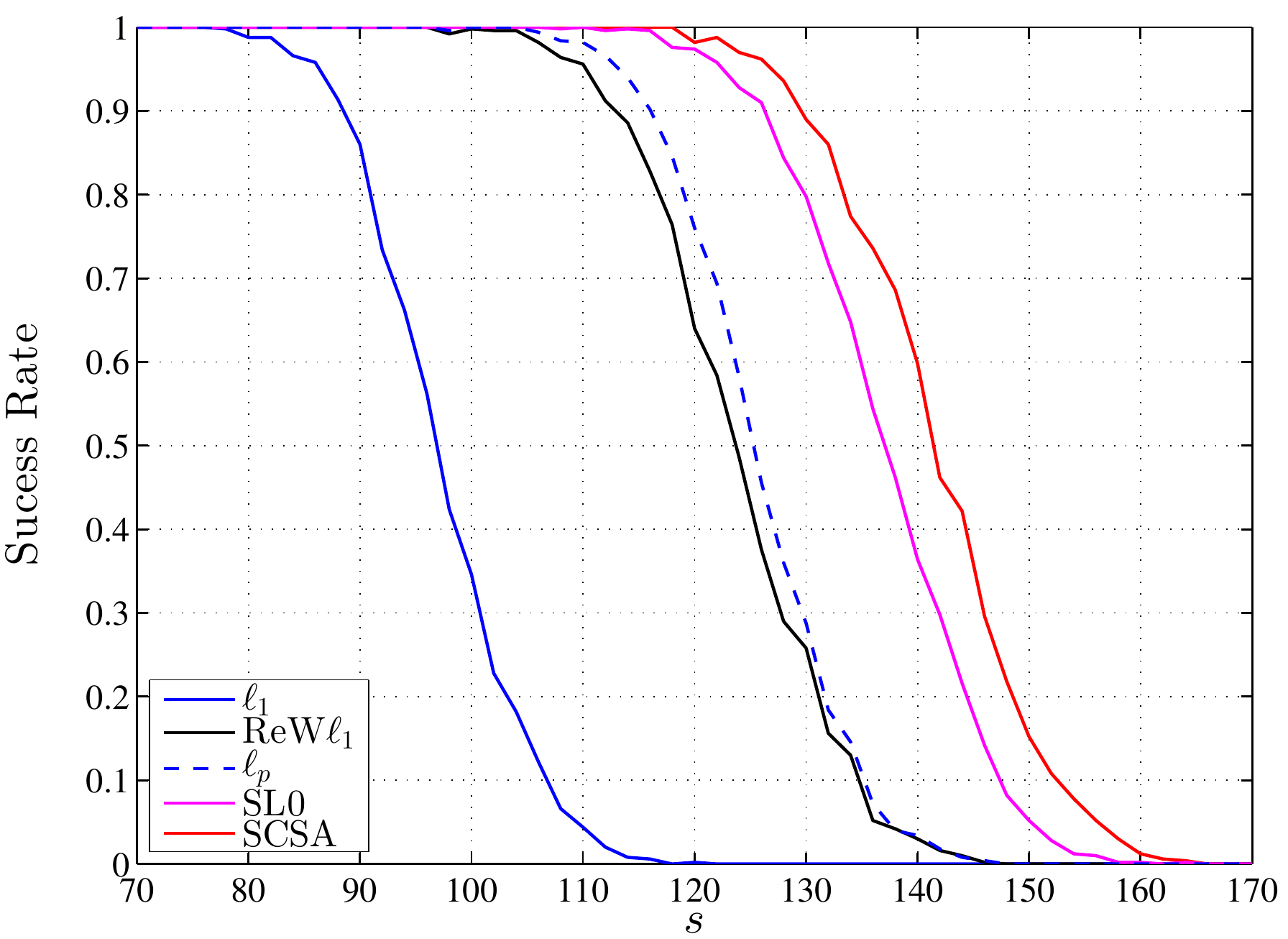}
        }%
        \vspace{-0.2cm}
        \subfigure{%
                \includegraphics[width=0.49\textwidth]{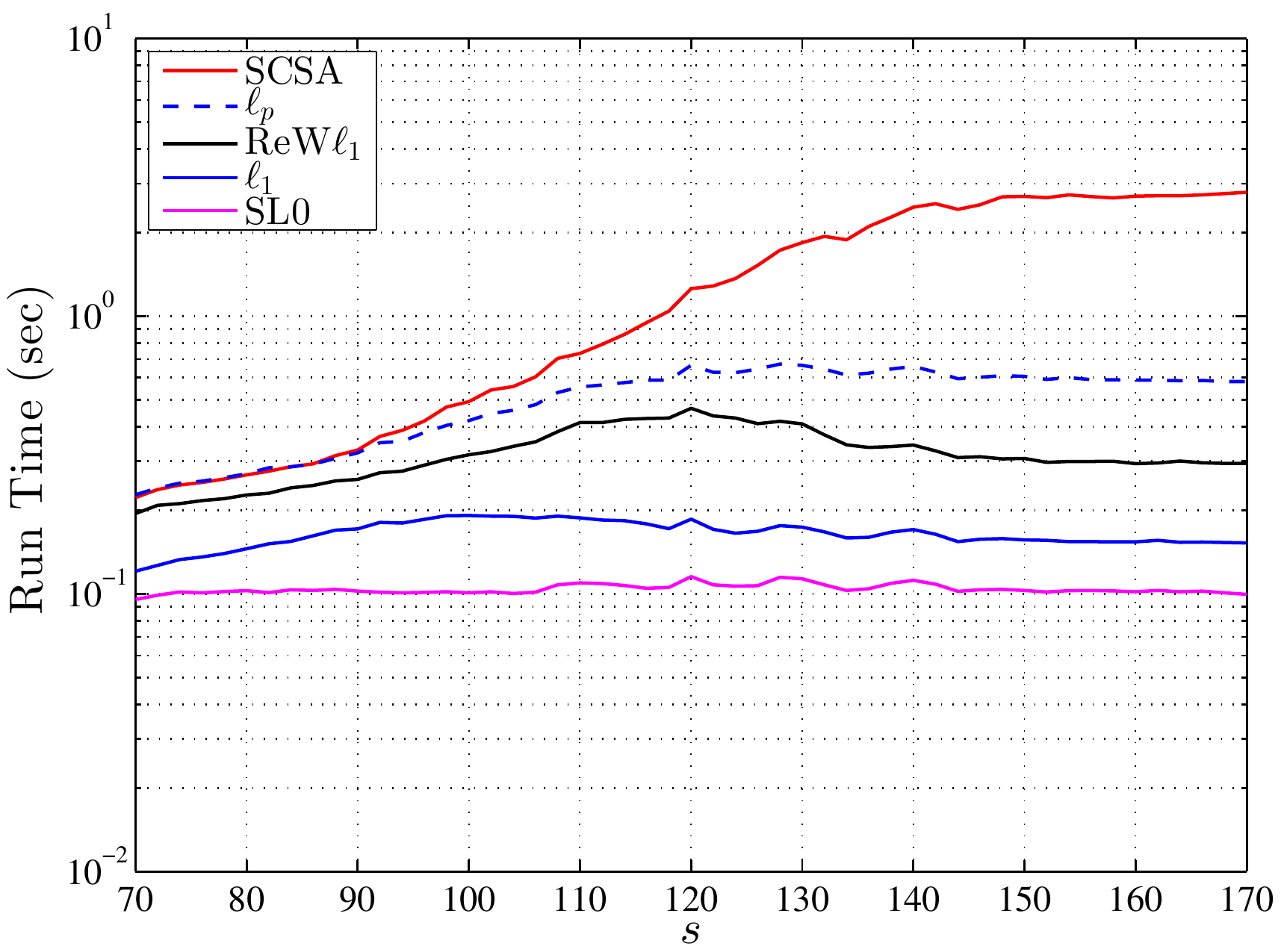}
        }%
        \vspace{-0.2cm}
        \caption{Comparison of the SCSA algorithm in solving the noise-free problem to $\ell_1$ minimization, the SL0 algorithm \cite{MohiBJ09}, $\ell_p$ quasi-norm minimization \cite{FoucL09,Char07} ($p=0.5$), and the reweighted $\ell_1$ minimization \cite{CandWB08} in terms of success rate and execution time. The dimensions of the sensing matrix are $250 \times 500$. Trials are repeated 500 times, and results are averaged over them.} \label{fig:NoNoise}
\end{figure}

The dimensions of the sensing matrix are fixed to $250 \times 500$, the sparsity of $\xbt$ is changed from 70 to 170, and success rate and average execution time over 500 trials are plotted in Fig.~\ref{fig:NoNoise}. As depicted in this figure, SCSA has the best success rate amongst the algorithms, whereas its computational load is much higher than the closest competitor. However, as we show shortly, in the noisy setting which is more realistic, SCSA maintains the superiority with a quite reasonable complexity.

\subsection{Recovery from Noisy Measurements} \label{sec:NoisyExp}
In the following experiments, superiority of the proposed algorithm in the noisy setting is demonstrated. Toward this end, the SCSA algorithm is compared with the oracle estimator \cite{CandT07}, which knows the location of the nonzero elements of the true solution, LASSO (or BPDN), {\color{\RTwo_Col}the SCAD penalty \cite{FanL01}}, Robust SL0 \cite{EfteBJM09}, the method of \cite{Char12,Char09}, and iterative log thresholding (ILT) \cite{MaliA13}. The description and implementation details of these algorithm are as follows.

To implement the LASSO estimator, IST and/or FISTA methods are used. {\color{\RTwo_Col} The smoothly clipped absolute deviation (SCAD) penalty is a well-known nonconvex function that promotes sparsity more tightly than the $\ell_1$ norm does and has some oracular properties \cite{FanL01}. For this penalty, we set the parameter $a$ to 3.7 as suggested in \cite{FanL01}. To efficiently solve the optimization problem resulting from the SCAD penalty, we use the algorithm of \cite{GassRC09} with parameter $\tau = 10^{-3}$ and the external loop stopping threshold equal to $10^{-4}$.} \footnote{{\color{\RTwo_Col}MATLAB code: https://sites.google.com/site/alainrakotomamonjy/}} Robust SL0 (RSL0), a modification to the original SL0 to handle noisy measurements, 
instead of a regularization parameter, needs the noise variance in the denoising step. To have a fair comparison, $\sigma_w$ is passed to it. Other parameters of RSL0 are the same as in the Experiment 2 except for \texttt{sigma\_min=}$\sigma_w /10$. The method of \cite{Char12,Char09}, which we refer to as IST-$p$, uses a generalized version of the soft thresholding operator, $\mathcal{S}_{\alpha}$, defined as
\begin{equation*}
\mathcal{S}_{\alpha}^{(p)}(x) = \max(|x|-\alpha|x|^{p-1},0) \sign(x);
\end{equation*}
otherwise, it is identical to the IST. We use two instances of this method with $p = 0.5$ and $p = 0.1$. The ILT algorithm is an extension of the reweighted $\ell_1$ minimization in \cite{CandWB08} to the noisy case. In fact, it solves the following optimization problem
\begin{equation*}
\min_{\xb} \lambda \summ \log(|x_i| + \beta) +  \| \Ab\xb - \bb\|^2,
\end{equation*}
in which $\beta$ is some small constant to ensure positivity of the argument of $\log(\cdot)$, using iterative thresholding approach.

For SCSA-FIT, SCSA-IT, IST, and FISTA, the regularization parameter is chosen according to the formula given in \eqref{LASSOlam}. For ILT{\color{\RTwo_Col}, SCAD,} and IST-$p$, the regularization parameter is numerically tuned at $\sigma_w = 10^{-2}$ and is linearly scaled with the change of noise standard deviation. The stopping criterion for IST, FISTA, IST-$p$, and ILT is $d = \| \xb_{i}  - \xb_{i-1} \| / \| \xb_{i-1} \| \leq \eta$, where $\xb_{i}$ and $\xb_{i-1}$ are the solutions at the $i$th and $(i-1)$th iterations, $\eta = \min(10^{-3} \lambda, 10^{-4})$, and $\lambda$ is the associated regularization parameter. For SCSA-FIT and SCSA-IT, $\epsilon_1$ and $\epsilon_2$ are set as suggested in Remark 2. 
Moreover, for IST, FISTA, ILT, and IST-$p$, $\mu$ is always fixed to $0.99/(2 \lambda_{max}(\Ab^T \Ab))$, while for SCSA-FIT and SCSA-IT, $\mu$ is $0.99/(2 \lambda_{max}(\Ab^T \Ab) + \lambda / \sigma)$.

To obtain accurate and stable results, similar to \cite{BenEE10}, $\MSNR$ is used in Experiments 3 and 4 to compare the performance of the algorithms. In fact, with $\MSNR$, we are comparing the `typical' performance of these algorithms as most performance guarantees for the LASSO estimator and other nonconvex estimators hold with some probability (see e.g., \cite{BibkRA09,BenEE10,Zhan10}).

\begin{figure}[tb]
        \centering
        \subfigure{%
                \includegraphics[width=0.49\textwidth]{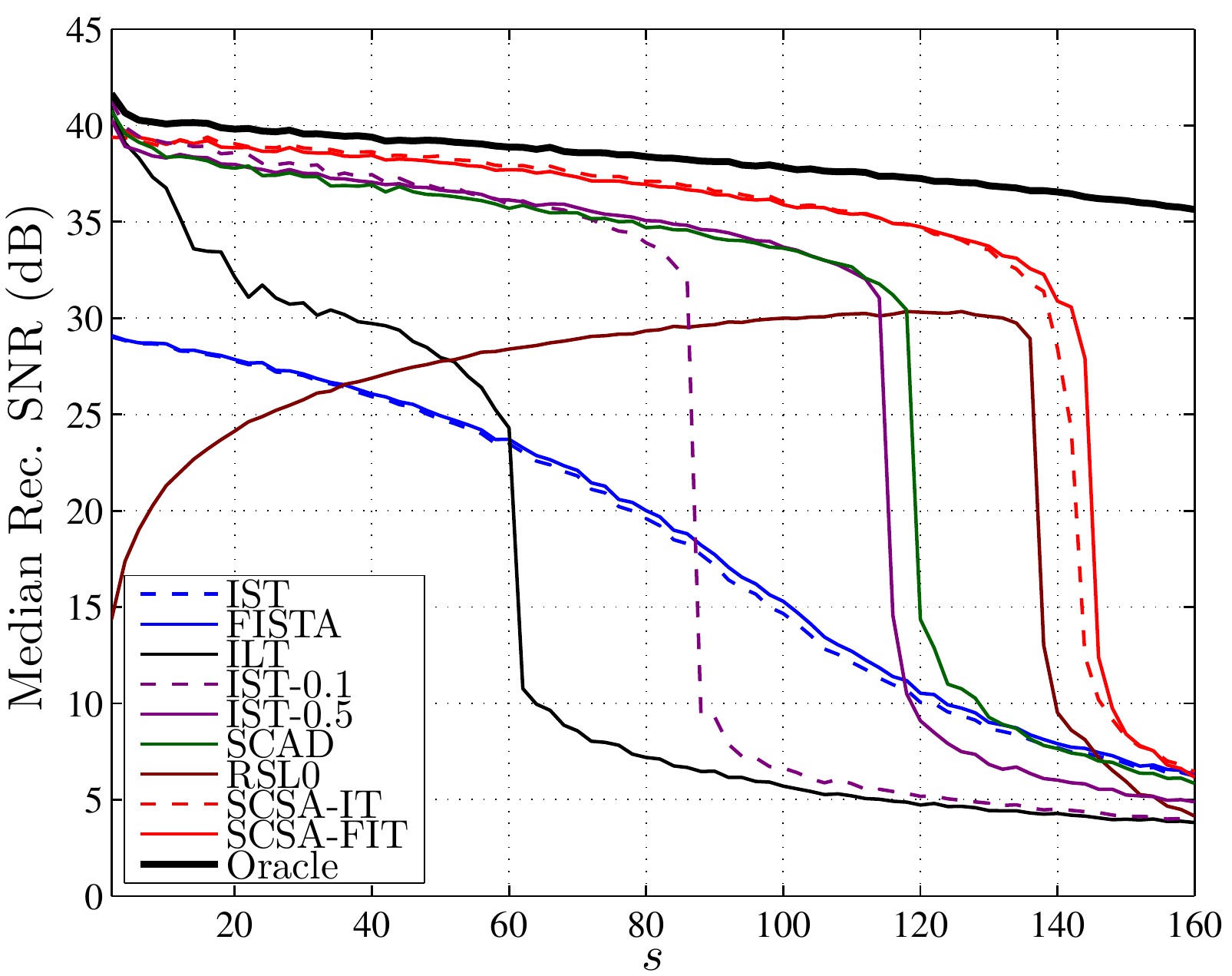}
        }%
        \vspace{-0.2cm}
        \subfigure{%
                \includegraphics[width=0.49\textwidth]{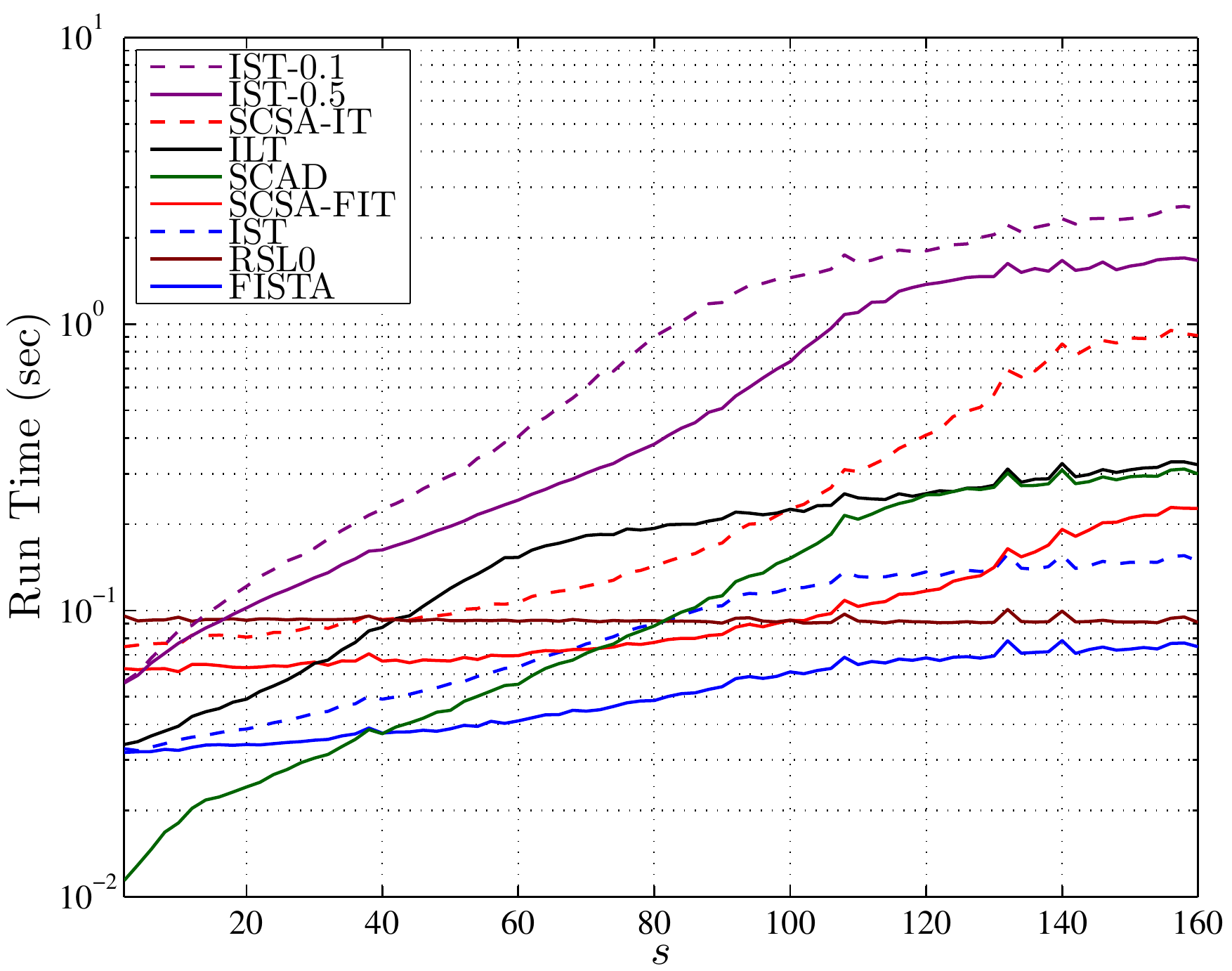}
        }%
        \vspace{-0.2cm}
        \caption{Comparison of two variants of the SCSA algorithm in solving the noisy problem to IST, FISTA, {\color{\RTwo_Col} SCAD \cite{FanL01},} ILT \cite{MaliA13}, RSL0 \cite{EfteBJM09}, IST-$p$ \cite{Char12,Char09}, and the oracle estimator \cite{CandT07} in terms of $\MSNR$ and execution time. The dimensions of the sensing matrix are $250 \times 500$. The sparse vectors are Gaussian distributed, and $\sigma_w = 10^{-2}$. Trials are repeated 500 times, and results are averaged over them.} \label{fig:GausComp}
\end{figure}

\emph{Experiment 3.} Under the above conditions, for Gaussian distributed sparse vectors, $s$ is changed from 2 to 160, and the $\MSNR$ and the averaged execution time (except for the oracle estimator) for all the algorithms over 500 runs are plotted in Fig.~\ref{fig:GausComp}. As clearly demonstrated in this figure, the (median) reconstruction SNR of the SCSA-FIT algorithm, for almost all values of $s$, is higher than others. Also, it has a near-oracle performance for a broader range of sparsity levels. So far as the computational load is concerned, SCSA-FIT needs at most (approximately) 3 times higher execution time in comparison to FISTA, the fastest algorithm for {\color{\RTwo_Col} most of sparsity levels}. However, the computational cost is {\color{\RTwo_Col} lower than that of SCAD and RSL0, when $s$ is larger than 70 and is smaller than 100, respectively. These two algorithms (SCAD and RSL0) are somehow the best competitors; however, they are not able to follow the oracular performance at the sparsity level that SCSA does.}

SCSA-IT and SCSA-FIT have a quite similar performance in terms of $\MSNR$. However, as expected, the former spends considerably more time than the latter to output a solution. Particularly, SCSA-FIT is approximately 8 times faster than SCSA-IT, when sparsity level is equal to 140.


\emph{Experiment 4.} To show that the SCSA algorithm is effective for sparse vectors which are not Gaussian-like distributed, we repeat Experiment 3 with the Rademacher distributed signals. The Rademacher distributed sparse vectors do not exhibit the power-law decaying behavior \cite{CandR05} when nonzero components are sorted according to their magnitude. 
In addition, some numerical simulations show that SL0, which parallels the idea of SCSA, does not work well for this kind of distributions (see nuit-blanche.blogspot.com/2011/11/post-peer-review-of-sl0.html).

\begin{figure}[tb!]
        \centering
        \subfigure{%
                \includegraphics[width=0.49\textwidth]{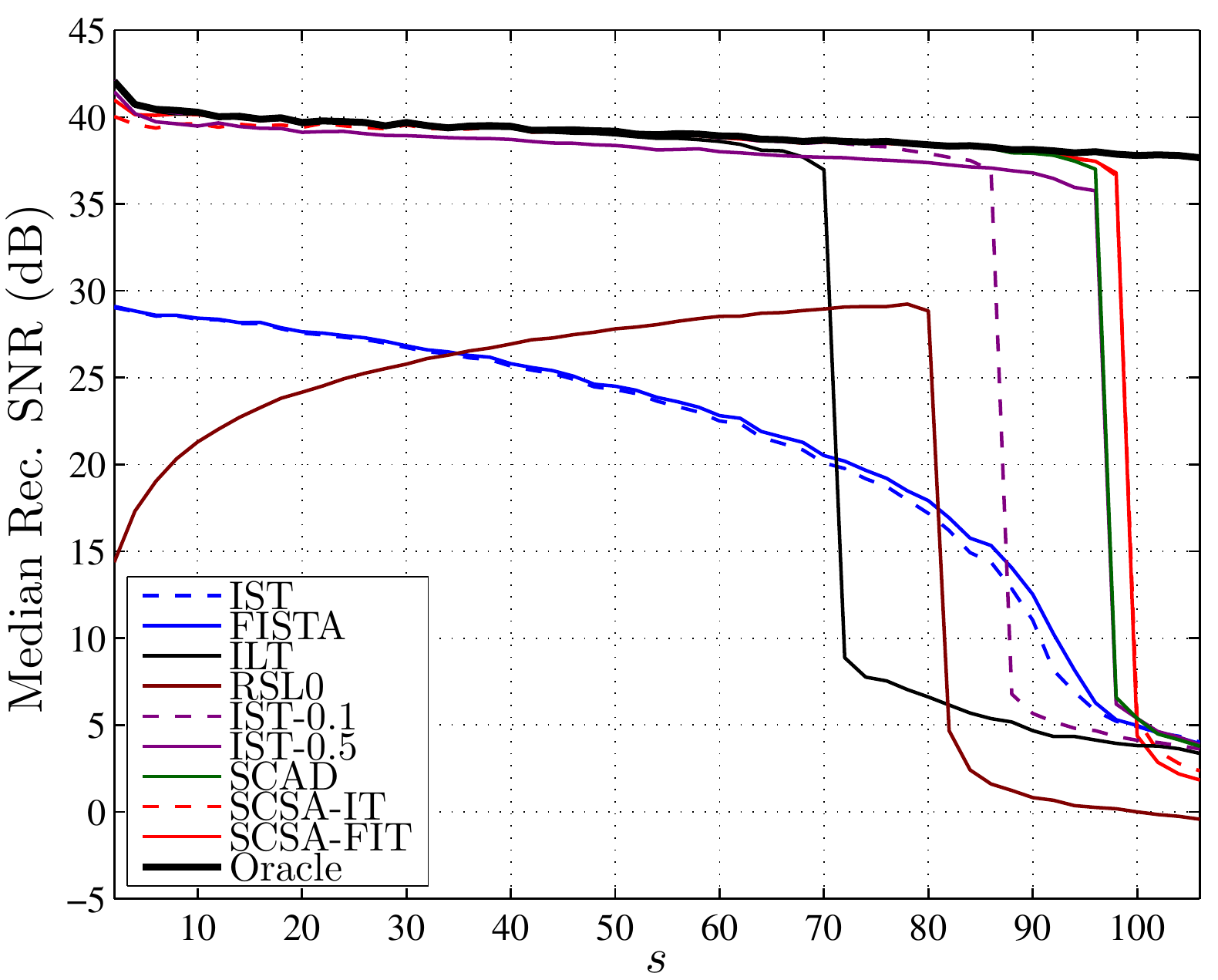}
        }%
        \vspace{-0.2cm}
        \subfigure{%
                \includegraphics[width=0.49\textwidth]{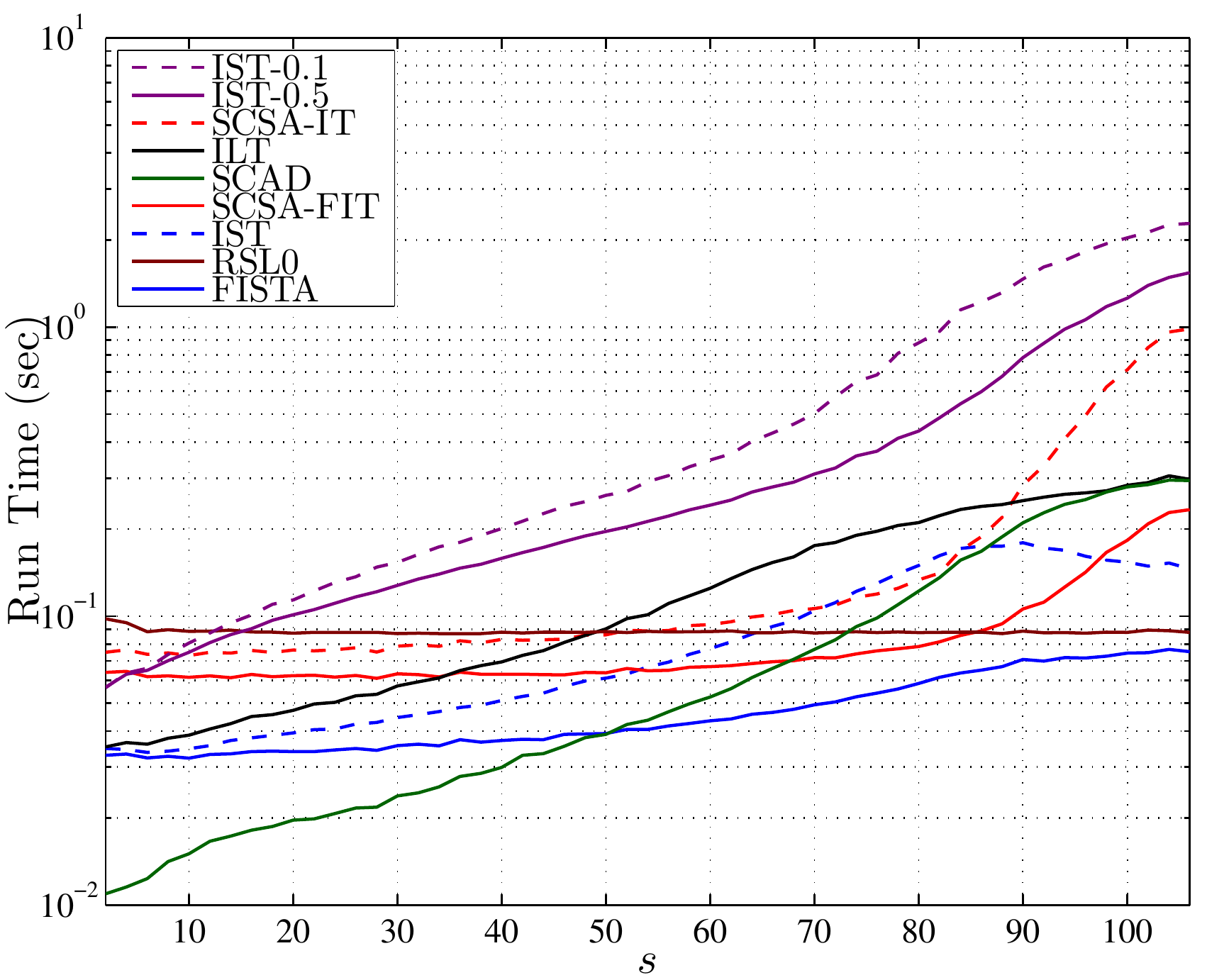}
        }%
        \vspace{-0.2cm}
        \subfigure{%
                \includegraphics[width=0.49\textwidth]{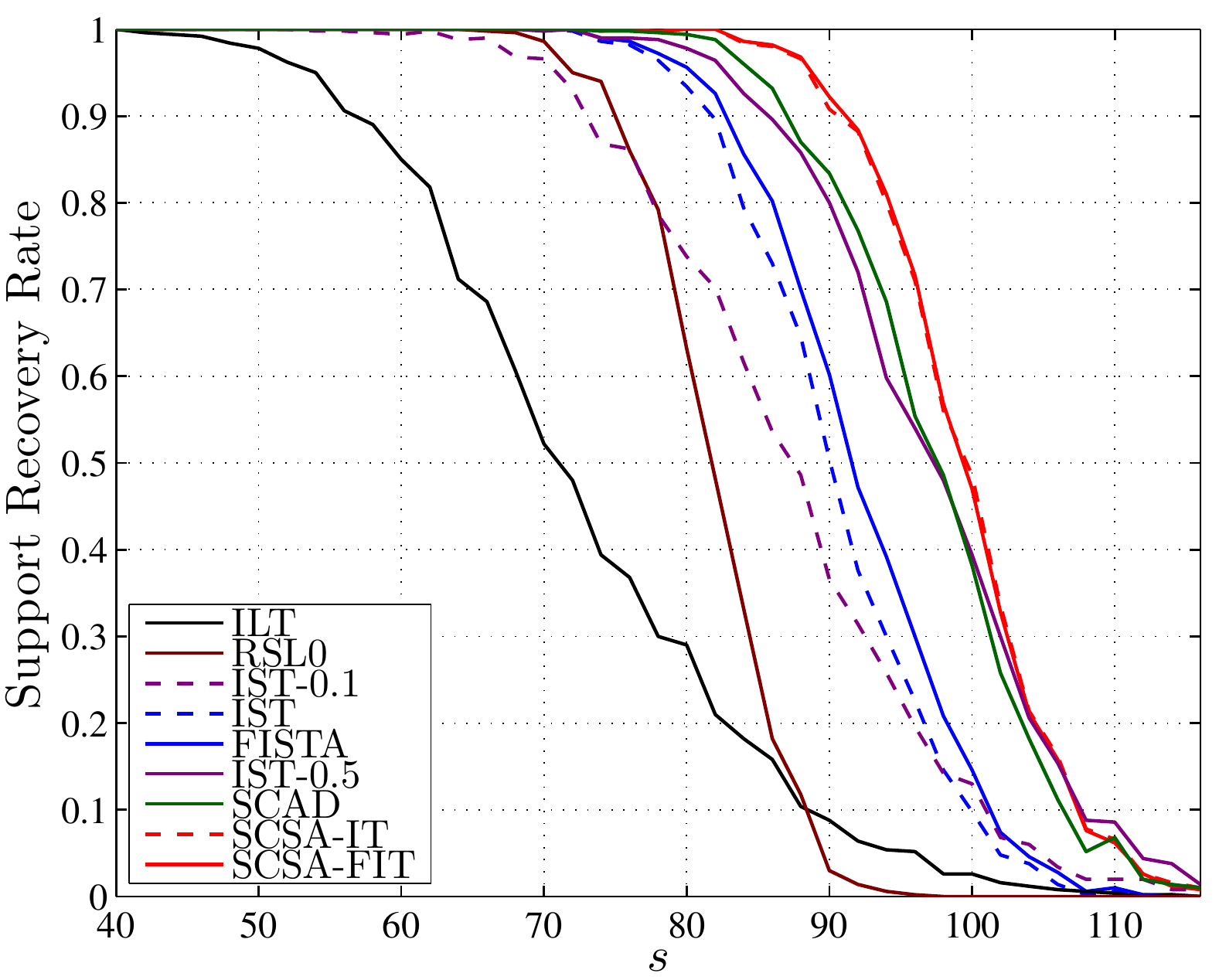}
        }%
        \vspace{-0.2cm}
        \caption{Comparison of two variants of the SCSA algorithm in solving the noisy problem with settings similar to those of Fig.~\ref{fig:GausComp} except that the sparse vectors are Rademacher distributed. Trials are repeated 500 times, and results are averaged over them. Since SCSA-FIT and SCSA-IT {\color{\RTwo_Col}as well as SCAD and IST-$0.5$} have very similar performances in terms of $\MSNR$, their traces are almost coincident in the top plot.}\label{fig:RadComp}
\end{figure}

{\color{\ROne_Col}In many applications, it is more crucial to find the support set of the sparse vector accurately \cite{Mill02}. To numerically asses the performance of SCSA in recovering the true support, we also calculate the SRR in this experiment.} Fig.~\ref{fig:RadComp} illustrates the results of the experiment for SCSA as well as all other algorithms in Experiment 3. As shown in this figure, SCSA and some other algorithms follow the oracle performance more closely than they did in Experiment 3. Furthermore, SCSA-FIT achieves the best performance in terms of $\MSNR$ {\color{\ROne_Col}and SRR}, whereas its complexity is quite comparable to FISTA.

\begin{figure*}[tb]
        \centering
        \subfigure[]{%
                \includegraphics[width=0.32\textwidth]{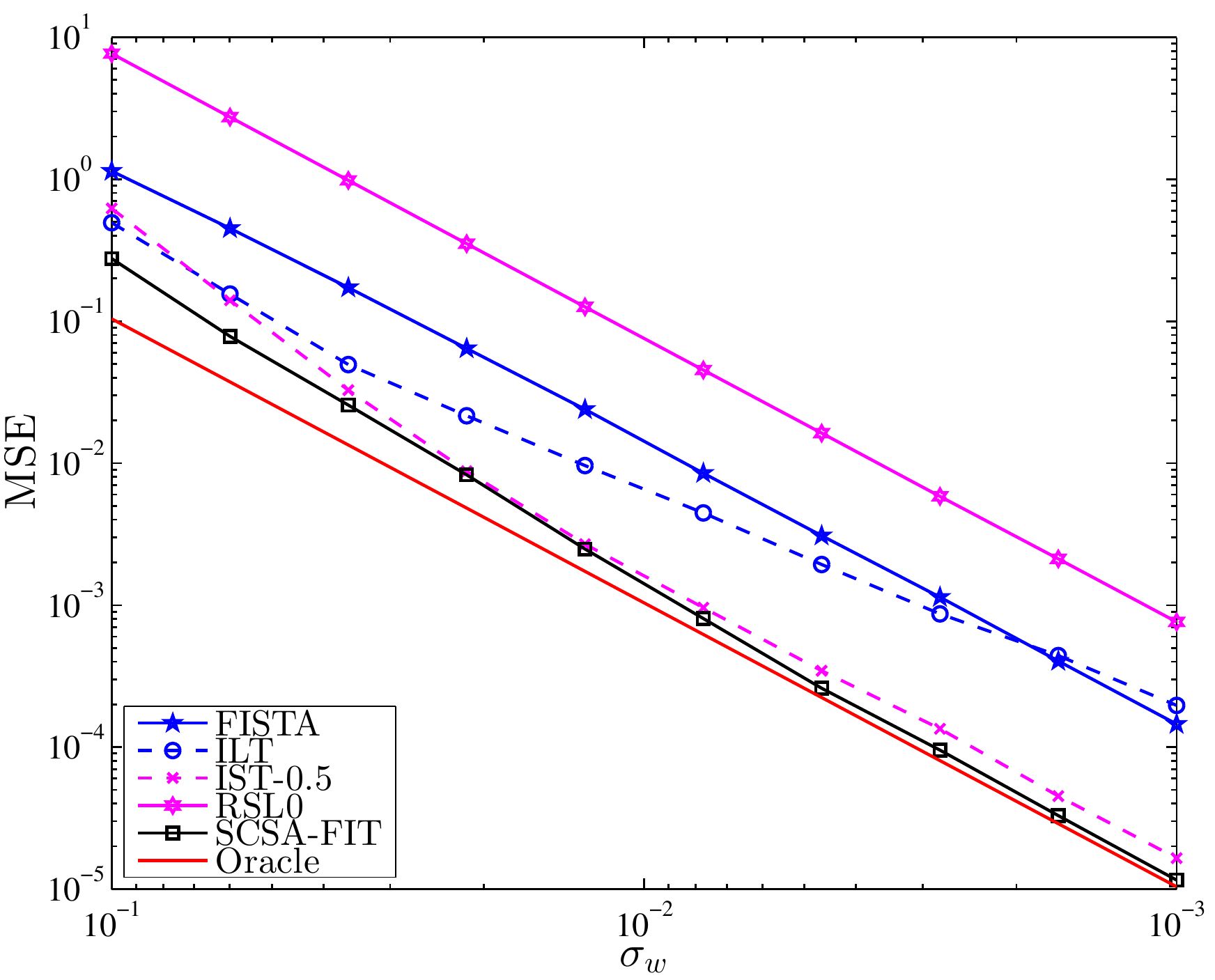}
        }%
        \subfigure[]{%
                \includegraphics[width=0.32\textwidth]{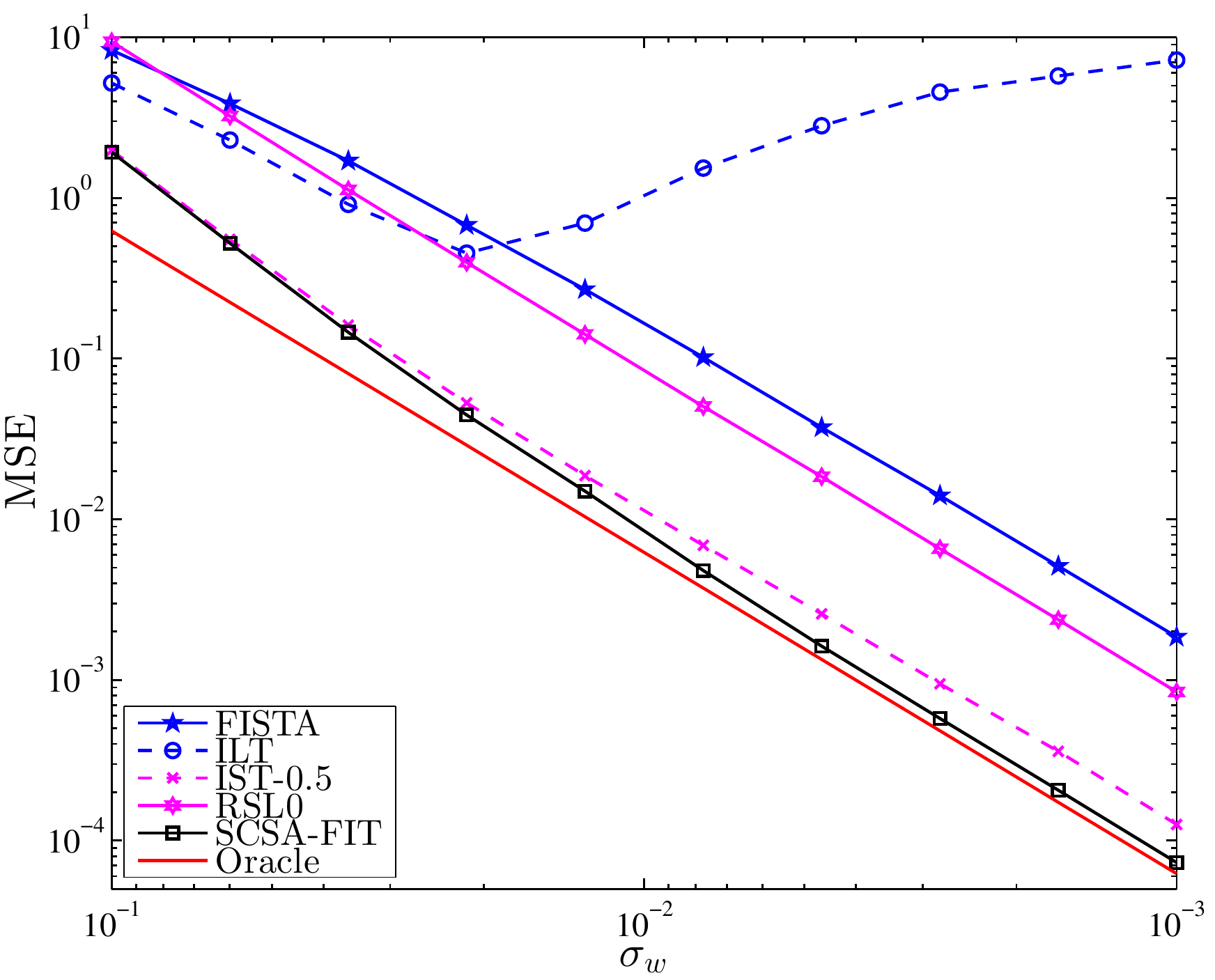}
        }%
        \subfigure[]{%
                \includegraphics[width=0.32\textwidth]{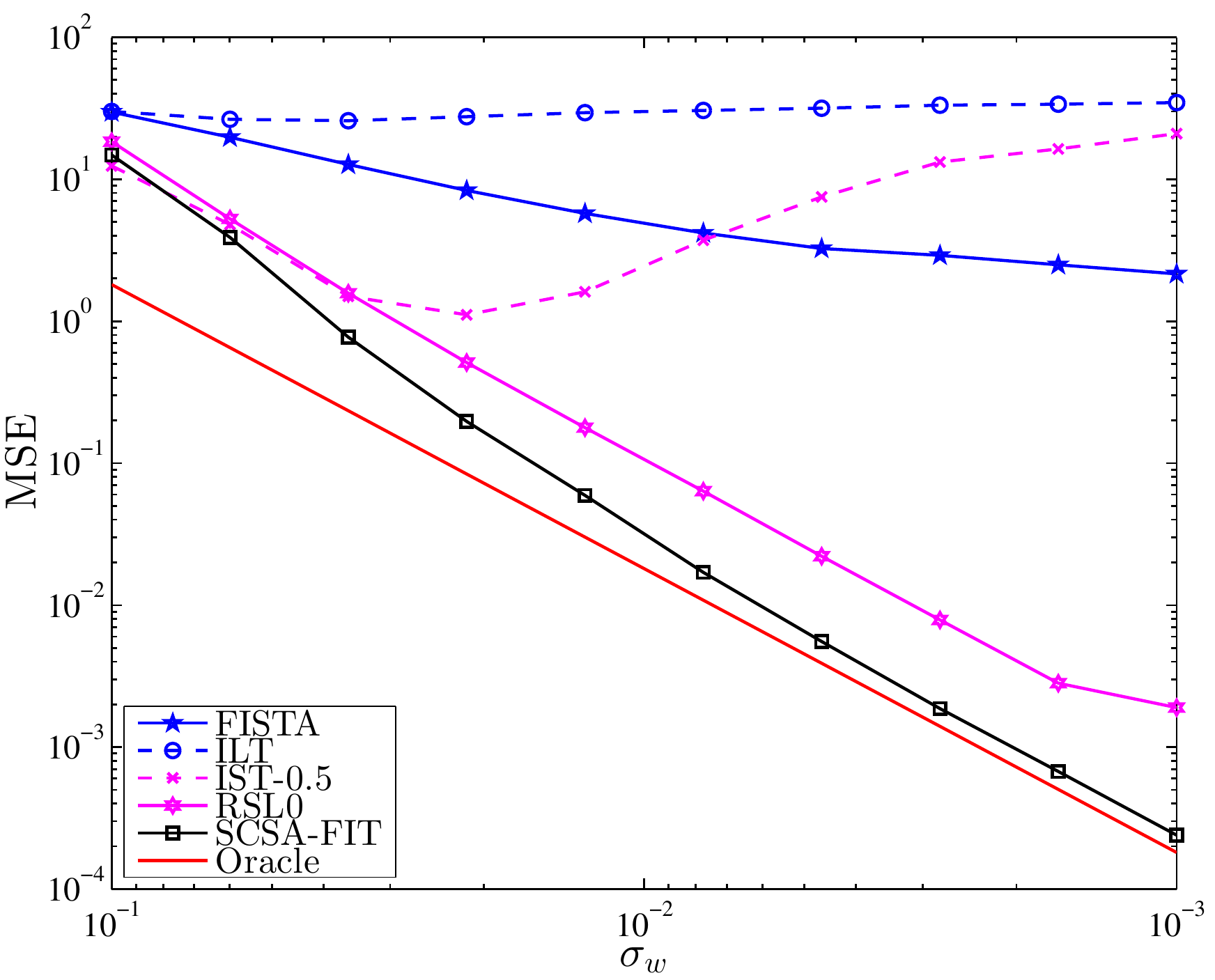}
        }%
        \vspace{-0.2cm}
        \caption{The mean-squared errors of SCSA-FIT, FISTA, ILT, IST-$0.5$, RSL0, and the oracle estimator in recovering Gaussian distributed sparse vectors from noisy measurements as a function of the noise standard deviation. The dimensions of the sensing matrix are $250 \times 500$. Trials are repeated 500 times, and results are averaged over them. (a), (b), and (c) correspond to $s=10,50,105$, respectively.}\label{fig:NoiseChange}
\end{figure*}

\emph{Experiment 5.} In this experiment, we examine and compare the accuracy of the SCSA-FIT algorithm in terms of MSE to FISTA, ILT, IST-$0.5$, SL0, and the oracle estimator as the noise variance changes. 
Under the same conditions as in Experiment 3 and for 3 sparsity levels ($s = 10, 50, 105$), the noise standard deviation is changed from $10^{-1}$ to $10^{-3}$, and the MSE is calculated. Results of this experiment are summarized in Fig.~\ref{fig:NoiseChange}. As clearly depicted, for all values of $s$, SCSA-FIT is the closest one to the oracle estimator, and can follow it even for the large sparsity level of $s=105$.

\section{Conclusion} \label{sec:Con}
For the problem of recovering sparse vectors from compressed measurements, we proposed to replace the $\ell_0$ norm with a family of concave functions that closely approximates sparsity. To solve the consequent nonconvex optimization problem, we exploited a continuation approach leading to starting from $\ell_1$ minimization, followed by successively solving accuracy-increasing approximations of the $\ell_0$ norm subject to the constraints. In the presence of noise in the measurements, we combined the continuation approach to iterative thresholding method of optimization and proposed a computationally inexpensive algorithm. This was obtained by deriving the closed-form solution for the program \eqref{FminIST2}. We also provided a choice for the regularization parameter in \eqref{FminUn} which is based on the regularization parameter for the LASSO estimator. The numerical simulations revealed that the proposed method considerably and consistently outperforms some of the common algorithms (including LASSO), especially when the sparsity level increases.

\appendices
\section{} \label{appShr}
Before deriving the closed-form solution to \eqref{FminIST2} for $\fsig(|x|) = 1 - \exp(-|x|/\sigma)$, we need to introduce the Lambert W function and prove an inequality on this function.

The Lambert W function, which has several applications in physics and applied and pure mathematics (particularly, combinatorics) \cite{CorlGHJK96}, is defined as the multivalued inverse of the function $w \mapsto w e^w$ \cite{CorlGHJK96}. More precisely, the Lambert W function, denoted by $W(x)$, is given implicitly by
\begin{equation*}
W(x) e^{W(x)} = x
\end{equation*}
where $x$ in general can be a complex number, but, herein, we only deal with real-valued $W$ and assume $x$ to be real. In this fashion, $W(x)$ is single-valued for $x \geq 0$ and is double-valued for $-\frac{1}{e} \leq x < 0$ \cite{CorlGHJK96}. To discriminate between the two branches for $x \in [-\frac{1}{e},0)$, we use the same notation as in \cite{CorlGHJK96} and denote the branch satisfying $W(x) \geq -1$ and $W(x) \leq -1$ by $W_0(x)$ and $W_{-1}(x)$, respectively.

\newtheorem{Lem1}{Lemma}
\begin{Lem1} \label{LemBranches}
For any $y \in [-\frac{1}{e},0)$, $W_0(y) + W_{-1}(y) \leq -2$.
\begin{proof}
Simple algebraic manipulation shows that the parabola $e^{-1}(\frac{1}{2}x^2 + x - \frac{1}{2})$ is below the function $xe^x$ for $x \in (-1,0)$ and above it for $x < -1$. Therefore, any line parallel to the $x$-axis with the $y$-intercept in the interval $y = [-\frac{1}{e},0)$, intersects the parabola at equal or larger $x$ coordinates than those corresponding to intersections with $xe^x$. From the other side, the $x$ coordinate of the intersection points of these lines with $xe^x$ gives $W_0(y)$ and $W_{-1}(y)$. Consequently, for any $y \in [-\frac{1}{e},0)$, the sum $W_0(y) + W_{-1}(y)$ is less than or equal to the sum of the roots of $e^{-1}(\frac{1}{2}x^2 + x - \frac{1}{2})=y$ which is always equal to $-2$. This completes the proof.
\end{proof}
\end{Lem1}

To obtain the solution to \eqref{FminIST2}, we first notice that since $\Fsig(\cdot)$ is a separable function, the minimizer can be obtained element by element. Therefore, we focus on the one-variable optimization problem. To that end, let $L(x,x_0) = \frac{1}{2\mu}(x - x_0)^2 + \lambda \fsig(|x|)$ denote the corresponding scalar version of \eqref{FminIST2} and let $x^* = \argmin_{x} L(x,x_0)$. It is easy to show that $L(x,-x_0) = L(-x,x_0)$; thus, from
\begin{equation*}
\left\{
\begin{IEEEeqnarraybox}[][c]{l?s}
\IEEEstrut
x^* = \argmin_{x} \{L(x,x_0)\} & for $x_0 \geq 0$ \\
x^* = - \argmin_{x} \{L(-x,x_0)\} & for $x_0 < 0$,
\IEEEstrut
\end{IEEEeqnarraybox}
\right.
\end{equation*}
we get
\begin{IEEEeqnarray*}{rCl}
x^* & = & \sign(x_0) \argmin_{x}\Big \{\frac{1}{2\mu}(\sign(x_0)x - x_0)^2 + \lambda \fsig(|x|) \Big\},\\
& = & \sign(x_0) \argmin_{x}\Big \{\frac{1}{2\mu}(x - |x_0|)^2 + \lambda \fsig(|x|) \Big\}.
\end{IEEEeqnarray*}
Let $y^* = \argmin_{x} \{\frac{1}{2\mu}(x - |x_0|)^2 + \lambda \fsig(|x|) \}$. This definition implies that $y^* \geq 0$, since if $y^* < 0$, then $\widehat{y} = -y^*$ contradicts the optimality of $y^*$ as $(\widehat{y} - |x_0|)^2 < (y^* - |x_0|)^2$. Consequently, we have
\begin{equation*}
x^* = \sign(x_0) \argmin_{x} \Big \{\frac{1}{2\mu}(x - |x_0|)^2 + \lambda \fsig(x)~|~x \geq 0 \Big \}.
\end{equation*}

Denoting $\widehat{L}(x,x_0) = \frac{1}{2\mu}(x - |x_0|)^2 + \lambda \fsig(x)$, putting $\fsig(x) = 1 - \exp(-x/\sigma)$, and differentiating $\widehat{L}(x,x_0)$ with respect to $x$, it can be obtained
\begin{equation*}
\frac{1}{\mu}(x - |x_0|) = -\frac{\lambda}{\sigma} e^{-\frac{x}{\sigma}} \Longrightarrow \frac{x - |x_0|}{\sigma} e^{\frac{x-|x_0|}{\sigma}} = - \frac{\mu \lambda}{\sigma^2} e^{- \frac{|x_0|}{\sigma}}.
\end{equation*}
The above equation admits two solutions
\begin{equation*}
\left\{
\begin{IEEEeqnarraybox}[][c]{l?s}
\IEEEstrut
x_1 = \sigma W_0(- \frac{\mu \lambda}{\sigma^2} e^{-\frac{|x_0|}{\sigma}}) + |x_0| \\
x_2 = \sigma W_{-1}(- \frac{\mu \lambda}{\sigma^2} e^{-\frac{|x_0|}{\sigma}}) + |x_0|.
\end{IEEEeqnarraybox}
\right.
\end{equation*}
Since, for $ x < -\frac{1}{e}$, $W(x)$ is not defined, we should have $- \frac{\mu \lambda}{\sigma^2} e^{-\frac{|x_0|}{\sigma}} \geq -\frac{1}{e}$ or $|x_0| \geq \sigma (1 + \ln (\mu \lambda) - 2 \ln(\sigma))$; otherwise, the minimizer of $\widehat{L}(x,x_0)$ lies at $x = 0$.

Substituting $x_1$ and $x_2$ in $\widehat{L}(x,x_0)$, letting $z = - \frac{\mu \lambda}{\sigma^2} e^{-\frac{|x_0|}{\sigma}}$, and using the definition $W(x) e^{W(x)} = x$, we have
\begin{IEEEeqnarray*}{rCl}
\widehat{L}(x_1,x_0) & = & \lambda + \frac{\sigma^2}{\mu} \Big( \frac{1}{2} W_0^2(z) + W_0(z) \Big), \\
\widehat{L}(x_2,x_0) & = & \lambda + \frac{\sigma^2}{\mu} \Big( \frac{1}{2} W_{-1}^2(z) + W_{-1}(z) \Big).
\end{IEEEeqnarray*}
From Lemma \ref{LemBranches}, we have
\begin{IEEEeqnarray*}{rCl}
& & \begin{cases}
W_0(z) + W_{-1}(z) \leq -2\\
W_0(z) \geq -1
\end{cases} \\
& & \quad \Rightarrow 0 \leq W_0(z) + 1 \leq -1 - W_{-1}(z) \\
& & \quad \Rightarrow \frac{1}{2} W_0^2(z) + W_0(z) \leq \frac{1}{2} W_{-1}^2(z) + W_{-1}(z).
\end{IEEEeqnarray*}
This shows that $x_2$ cannot be a minimizer of $\widehat{L}(x,x_0)$. In addition, it can be easily checked that $\widehat{L}(x,x_0)$ for $x \geq 0$ is convex only when $\sigma^2 \geq \mu \lambda$. Therefore, it can occur that the minimizer resides at the border ($x=0$). It is very hard to analytically find the condition under which $x=x_1$ or $x=0$ is the minimizer. However, one can easily compare the cost function at these points to find the minimizer. In summary, the one dimensional shrinkage operator can be characterized as
\begin{equation} \label{ShrDef}
\mathcal{T}_{\mu\lambda}^{(\sigma)}(x_0) =
\left\{
\begin{IEEEeqnarraybox}[][c]{l?s}
\IEEEstrut
0 & $|x_0| \geq \sigma (1 + \ln (\mu \lambda / \sigma^2))$\\
0 & $\widehat{L}(x_1,x_0) \geq \widehat{L}(0,x_0)$\\
\sigma W_0(z) + |x_0| & otherwise,
\end{IEEEeqnarraybox}
\right.
\end{equation}
where $z = - \frac{\mu \lambda}{\sigma^2} e^{-\frac{|x_0|}{\sigma}}$.

\section{Proof of Theorem \ref{LConvIT}} \label{ISTConv}
For the sake of simplicity, let us define $\phi(\xb) \triangleq \lambda_{\sigma} \Fsig(\xb)$, where, without introducing any ambiguity, we omitted the subscript $\sigma$. Further, let $g(\xb) \triangleq h(\xb) + \phi(|\xb|)$ denote the cost function in \eqref{FminUn}. The program \eqref{FminIST} now is equal to
\begin{equation} \label{FminISTShort}
\xb_{k+1} = \argmin_{\xb}\Big \{ \langle \xb - \xb_k, \nabla h(\xb_k) \rangle + \frac{1}{2\mu} \|\xb - \xb_k\|^2 + \phi(|\xb|) \Big \}.
\end{equation}
Since $h(\xb)$ has an $M$-Lipschitz continuous gradient, we have \cite{OrteR00}
\begin{equation} \label{hIneq}
h(\xb_{k+1}) \leq h(\xb_{k}) + \langle \xb_{k+1} - \xb_{k}, \nabla h(\xb_{k}) \rangle + \frac{M}{2} \| \xb_{k+1} - \xb_{k} \|^2.
\end{equation}
Moreover, $\nabla^{2}\phi(\xb) \succeq -M'\Ib$ for $\xb \geq 0$, implies that, for all $\xb, \yb \geq \Zerb$,
\begin{equation*}
\phi(\xb) \leq \phi(\yb) + \langle \xb - \yb, \nabla \phi(\xb)\rangle + \frac{M'}{2} \| \yb - \xb \|^2.
\end{equation*}
Substituting $\xb$ and $\yb$ with $|\xb_{k+1}|$ and $|\xb_{k}|$, respectively, we get
\begin{IEEEeqnarray*}{rCl}
\phi(|\xb_{k+1}|) & \leq & \phi(|\xb_{k}|) + \langle |\xb_{k+1}| - |\xb_{k}|, \nabla \phi(|\xb_{k+1}|)\rangle \\
& & ~+ \frac{M'}{2} \| |\xb_{k+1}| - |\xb_{k}| \|^2.
\end{IEEEeqnarray*}
Using $\| |\xb| - |\yb| \| \leq \| \xb - \yb \|$, the above inequality resorts to
\begin{IEEEeqnarray}{rCl} \label{fIneq}
\phi(|\xb_{k+1}|) & \leq & \phi(|\xb_{k}|) + \langle |\xb_{k+1}| - |\xb_{k}|, \nabla \phi(|\xb_{k+1}|) \rangle \nonumber \\
 &  & + \frac{M'}{2} \| \xb_{k+1} - \xb_{k} \|^2.
\end{IEEEeqnarray}
From the other side, $\xb_{k+1}$ is a minimizer of \eqref{FminISTShort}, so we can write
\begin{equation} \label{OptIneq}
\langle \xb_{k+1} - \xb_{k}, \nabla h(\xb_k) \rangle \leq \phi(|\xb_{k}|) - \phi(|\xb_{k+1}|) - \frac{1}{2\mu} \| \xb_{k+1} - \xb_{k} \|^2.
\end{equation}
First-order concavity condition for $\phi$ implies that $\phi(\yb) \leq \phi(\xb) + \langle \yb - \xb, \nabla \phi(\xb) \rangle$. Replacing $\xb$ and $\yb$ with $|\xb_{k+1}|$ and $|\xb_{k}|$, respectively, we get
\begin{equation} \label{ConIneq}
\phi(|\xb_{k}|) - \phi(|\xb_{k+1}|) + \langle |\xb_{k+1}| - |\xb_{k}|, \nabla \phi(|\xb_{k+1}|) \rangle \leq 0.
\end{equation}
Combining \eqref{hIneq} and \eqref{fIneq} results in
\begin{IEEEeqnarray}{rCl}
\IEEEeqnarraymulticol{3}{l}{g(\xb_{k+1}) - g(\xb_{k})} \nonumber \\ \quad
   & \leq & \langle \xb_{k+1} - \xb_{k}, \nabla h(\xb_{k}) \rangle + \langle |\xb_{k+1}| - |\xb_{k}|, \nabla \phi(|\xb_{k+1}|)\rangle \nonumber \\
   &  & +  \frac{1}{2} (M + M')\|\xb_{k+1} - \xb_{k}\|^2 \nonumber \\
& \overset{(a)}{\leq} & \phi(|\xb_{k}|) - \phi(|\xb_{k+1}|) + \langle |\xb_{k+1}| - |\xb_{k}|, \nabla \phi(|\xb_{k+1}|)\rangle \nonumber \\
&  & + \frac{1}{2} (M + M'-\frac{1}{\mu})\|\xb_{k+1} - \xb_{k}\|^2, \label{IneqT1}
\end{IEEEeqnarray}
where (a) follows from \eqref{OptIneq}. Assume that $0 < \mu < \frac{1}{M+M'}$, then \eqref{IneqT1} rearranges to
\begin{IEEEeqnarray}{rCl}
\IEEEeqnarraymulticol{3}{l}{\frac{1}{2} (\frac{1}{\mu} - M - M')\|\xb_{k+1} - \xb_{k}\|^2} \nonumber \\ \qquad \qquad \qquad \quad
 & \leq & g(\xb_{k}) - g(\xb_{k+1}) + \phi(|\xb_{k}|) - \phi(|\xb_{k+1}|) \nonumber \\
 & & + \langle |\xb_{k+1}| - |\xb_{k}|, \nabla \phi(|\xb_{k+1}|) \nonumber \\
& \overset{(b)}{\leq} &  g(\xb_{k}) - g(\xb_{k+1}), \label{IneqT2}
\end{IEEEeqnarray}
where (b) follows from \eqref{ConIneq}.

Now, we show that $\{g(\xb_k)\}$ is a nonincreasing sequence. Similar to \eqref{hIneq}, one has
\begin{equation*}
h(\xb) \leq h(\xb_{k}) + \langle \xb - \xb_{k}, \nabla h(\xb_{k})\rangle + \frac{M}{2} \| \xb - \xb_{k} \|^2.
\end{equation*}
As a result, if $\mu \in (0,\frac{1}{M})$, then $h(\xb) \leq H(\xb,\xb_k) \triangleq h(\xb_k) + \langle \xb - \xb_k , \nabla h(\xb_k) \rangle + \frac{1}{2 \mu} \| \xb - \xb_k \|^2$, or, equivalently,
\begin{equation*}
g(\xb) \leq H(\xb,\xb_k) + \phi(|\xb|) ~~ \forall \xb.
\end{equation*}
Replacing $\xb$ with $\xb_{k+1}$, we get $g(\xb_{k+1}) \leq H(\xb_{k+1},\xb_k) + \phi(|\xb_{k+1}|)$. Furthermore, $\xb_{k+1}$ is a solution to \eqref{FminISTShort}, so $H(\xb_{k+1},\xb_k) + \phi(|\xb_{k+1}|) \leq H(\xb_{k},\xb_{k}) + \phi(|\xb_{k}|) = g(\xb_k)$. This inequality together with the previous one leads to
\begin{equation*}
g(\xb_{k+1}) \leq g(\xb_{k}) ~~ \forall k \geq 0.
\end{equation*}
Summing \eqref{IneqT2} over $k=0,\cdots,N$, we get
\begin{equation*}
\frac{1}{2}(\frac{1}{\mu}-M-M') \sum_{k=0}^{N} \|\xb_{k+1}-\xb_{k}\|^2 \leq g(\xb_0) - g(\xb_{N+1}).
\end{equation*}
Since $g(\xb_0)$ is finite, we conclude that $\{\xb_k\}$ is convergent. Next, we prove that $\{\xb_k\}$ converges to a stationary point of \eqref{FminUn}.
Suppose that $\{\xb_k\}$ converges to $\xb^*$. It can be verified that the cost function in \eqref{FminISTShort} is strictly convex for $\mu < 1/M$; thus, the minimizer of \eqref{FminISTShort} is unique. This fact together with \cite[Thm. 1.17 and Thm. 7.41]{Rock98} implies that when $k \to \infty$, we have
\begin{equation*}
\xb^* = \argmin_{\xb}\Big\{ \langle \xb - \xb^*, \nabla h(\xb^*) \rangle + \frac{1}{2 \mu} \|\xb - \xb^*\|^2 + \phi(|\xb|) \Big\}.
\end{equation*}
First-order optimality condition of the above program leads to
\begin{equation*}
\Zerb \in \nabla h(\xb^*) + \frac{1}{\mu} (\xb - \xb^*) + \partial \phi(|\xb|) \quad \text{at} ~ \xb = \xb^*,
\end{equation*}
proving that $\xb^*$ is a stationary point of \eqref{FminUn}. \hfill \QEDclosed

\bibliographystyle{IEEEbib}
\bibliography{IEEEabrv,SCSA}

\end{document}